\documentclass[prc,aps,superscriptaddress,showpacs,nofootinbib]{revtex4}
\usepackage{amssymb}
\usepackage[dvips]{graphicx}
\usepackage[english]{babel}
\usepackage{indentfirst}
\usepackage{amsxtra}
\usepackage{amsmath}
\usepackage{supertabular}
\usepackage{multirow}
\usepackage[mathcal]{eucal}
\usepackage[usenames]{color}
\usepackage{ulem}

\begin{document}

\title{\textbf{Second--order equation of state with the Skyrme interaction.
Cutoff and dimensional regularization with the inclusion of rearrangement terms }}
\author{C.J. Yang}
\affiliation{Institut de Physique Nucl\'eaire, CNRS-IN2P3, Universit\'e Paris-Sud,
Universit\'e Paris-Saclay, 91406 Orsay, France}
\author{M. Grasso}
\affiliation{Institut de Physique Nucl\'eaire, CNRS-IN2P3, Universit\'e Paris-Sud,
Universit\'e Paris-Saclay, 91406 Orsay, France}
\author{X. Roca-Maza}
\affiliation{Dipartimento di Fisica, Universit\`a degli Studi di Milano, Via Celoria 16,
20133 Milano, Italy}
\affiliation{INFN, Sezione di Milano, Via Celoria 16, 20133 Milano, Italy}
\author{G. Col\`o}
\affiliation{Dipartimento di Fisica, Universit\`a degli Studi di Milano, Via Celoria 16,
20133 Milano, Italy}
\affiliation{INFN, Sezione di Milano, Via Celoria 16, 20133 Milano, Italy}
\author{K. Moghrabi}
\affiliation{Faculty of Sciences, Lebanese University, Beirut, Lebanon}
\affiliation{American University of Science and Technology, Beirut, Lebanon}

\begin{abstract}
We evaluate the second--order (beyond--mean--field) contribution to the equation of state of nuclear matter with the effective Skyrme force and use 
cutoff and dimensional regularizations to treat the ultraviolet divergence produced by the zero--range character of this interaction. 
An adjustment of the force parameters is then performed in both cases to remove any double counting 
generated by the explicit computation of beyond--mean--field corrections with the Skyrme force. In addition, we include at second order the rearrangement terms associated to the density--dependent part of the Skyrme force and discuss their effect.
Sets of parameters are proposed to define new effective forces which are specially designed for second--order calculations in nuclear matter. 
\end{abstract}

\maketitle

\vskip 0.5cm


\section{Introduction}

The energy--density--functional (EDF) theory was developed in nuclear physics in the last
decades. In this theoretical framework, the energy of the system is computed by
using functionals of the density, which are usually derived from effective
interactions \cite{bender}, with some exceptions (see, for instance, the
work reported in Ref. \cite{korte}). Mean--field--type models constitute the
 basis on which this theory was constructed starting from the
70s. Such models represent have strong analogies with the leading order of the many--body Dyson perturbative 
expansion \cite{FW}, based on the independent--particle approximation, and are 
currently applied to numerous many--particle systems.

Mean--field--based models are extensively employed in the study of
medium--mass and heavy nuclei, and are particularly successful in describing
with a good accuracy a large number of known masses and radii. Despite this
success, more sophisticated beyond--mean--field models are necessary, for
instance, to perform accurate spectroscopic analyses for nuclear ground
states or to provide reliable descriptions of the physical fragmentation of
nuclear excitations. In these cases, additional correlations, with respect
to what is contained in a mean--field picture, have to be included in the
theoretical scheme. Going beyond the mean field, with respect to the
many--body Dyson perturbative expansion, implies including higher orders.
This is a challenging task to accomplish for several reasons: among them,
the highly increased numerical cost, and the conceptual problems implied by
the choice of the interaction to be used. Some of these problems 
are mentioned in what follows. Within the EDF theory, the
currently used density functionals are usually produced by phenomenological
interactions or Lagrangians in the non relativistic and relativistic cases,
respectively. Skyrme \cite{skyrme,skyrme1} and Gogny \cite{gogny1,gogny2}
forces are the most used interactions in the non relativistic framework. The
fitting procedure of their parameters is performed with mean--field
calculations for nuclear matter and some chosen nuclei. It is obvious
that, if such interactions are used in cases where higher orders are
explicitly included in the theoretical models, double--counting problems
arise. Furthermore, in those cases where the effective interactions have a
zero range, ultraviolet divergences may occur beyond the mean field and have
to be treated.

Our objective is to construct a generalized EDF framework, where new
effective interactions are introduced, which are now designed to treat both
matter and finite nuclei in beyond--mean--field models, avoiding
double--counting problems and regularizing ultraviolet divergences. We
started this work with the Skyrme interaction \cite%
{moghraprl,moghra2012-1,moghra2012-2}. Second--order calculations were
performed to compute the equation of state (EOS) of symmetric, neutron, and
asymmetric matter. First, only symmetric matter and a simplified form of the
Skyrme interaction (contact interaction with a density--dependent coupling
constant) were analyzed \cite{moghraprl}. This simplified Skyrme model
corresponds to the so--called $t_{0}-t_{3}$ model, where the
velocity--dependent, the spin--orbit, and the tensor terms of the Skyrme
force are omitted. The second--order contribution was evaluated analytically
and its cutoff--dependent part was identified. The divergent part was found
to have a linear asymptotic behavior with respect to a momentum cutoff. The
set of parameters (three in that case) were adjusted on a benchmark EOS for
several values of the cutoff. These parameter sets were recently employed in
a simplified test--calculation for the nucleus $^{16}$O \cite{brenna}. The
second--order correction to the total binding energy was evaluated in this
nucleus and encouraging results were found indicating a reasonable
convergence with respect to the chosen cutoff.

We then performed the same type of procedure (analytical derivation of the
second--order contribution and adjustment of the parameters) by including
also the velocity--dependent terms in the Skyrme interaction. Two directions
were explored, namely: (i) keeping the cutoff--dependent terms \cite{moghra2012-1}, and (ii) applying the dimensional regularization technique
to extract only the finite part \cite{moghra2012-2}.

We have recently realized that some aspects of the formal derivation on
which both studies \cite{moghra2012-1} and \cite{moghra2012-2} were based
are not correct and we aim here at providing the correct full formulae and
results.
For the sake of clarity and to present a self-contained reference on the topic, we provide here all the needed details referring previous literature, new formulae and, for an easy comparison with Refs.~\cite{moghra2012-1,moghra2012-2}, the same type of figures. 
In addition, we illustrate the effect on the EOS related to the inclusion of the proper rearrangement terms at second order. 
Rearrangement terms were neglected in Refs. \cite{moghra2012-1,moghra2012-2}. The general way of computing 
them in beyond--mean--field models was discussed in Ref. \cite{waro}.

The ultraviolet divergence at second order is handled
in two ways: (i) by using a cutoff regularization, as in Ref. \cite%
{moghra2012-1}. The cutoff--dependent results are adjusted on a benchmark
EOS and several sets of parameters are generated for each chosen value of
the cutoff; (ii) by applying the dimensional--regularization procedure. Only
the finite parts of the second--order corrections are thus extracted as in
Ref. \cite{moghra2012-2} and the cutoff--dependent part is eliminated by
the dimensional regularization. A unique set of parameters is thus produced by
adjusting on a benchmark EOS.

As a first approximation, in the first part of the article we replace the effective mass $m^*$ with the bare mass 
$m$, as done in Ref. \cite{kaiser} (this approximation will be abandoned in the last part of the article).
We  find in this case that the finite part of the EOS calculated for symmetric and
neutron matter is coherent with the recent results shown in Ref. \cite%
{kaiser}. Differently from Ref. \cite{kaiser}: (1) tensor and spin--orbit
parts are omitted for simplicity in the interaction. For the tensor part,
this is first justified by the fact that many currently used Skyrme
interactions do not contain such terms. Spin--orbit and tensor terms do not
contribute to the first--order EOS. In principle, they contribute when the
second order is included; \footnote{%
We note that such terms were taken into account in Ref. \cite{kaiser};
however, the adjustment of the parameters performed in that work for the
second--order EOS of symmetric matter (done by omitting the
density--dependent term) provided $W_{0}=t_4=t_5=0$, where $W_{0}$, $t_4$,
and $t_5$ are the parameters that tune the spin--orbit ($W_0$) and tensor ($%
t_4$ and $t_5$) terms introduced in Ref. \cite{kaiser}.} (2) we also
calculate the cutoff--dependent part in the second--order contribution
because we perform a cutoff regularization; (3) also asymmetric matter is
treated here; (4) the density--dependent part of the Skyrme interaction is
not omitted. It is well known that this term is necessary to reproduce the
correct saturation point of symmetric matter at first order. In Ref. \cite%
{kaiser} it is stressed that this term is indeed necessary for properly
reproducing the saturation density of symmetric matter also at second order.
The second--order curves obtained there by omitting the density--dependent
term do not reproduce at all the saturation point providing a saturation
density of $\sim $ 0.22 fm$^{-3}$. It is well known 
that the inclusion of such term may be
problematic in many respects (see, for instance, Ref. \cite{stringari}).
Several drawbacks related to the density dependence were identified such
as, for instance, the existence of pathologies in some applications of 
the generator--coordinate method \cite{lac}. 
Our pragmatic
choice is to use density--dependent interactions as a starting point in our
work because of their good performance, and we leave for a future work 
the discussion of the associated
problems. 
Note that, despite the
explicit density dependence in the interaction, the Hugenholtz-van Howe
theorem \cite{huge} is satisfied in the Skyrme case whenever the proper rearrangement terms (associated to the density dependence) are
explicitly introduced, as is usually done within the random--phase approximation (RPA) and as 
was recently done also in the specific case of the second RPA \cite{rearra}.  Reference \cite{car} pointed   
out that the matrix elements of the interaction used for the computation of the energy in second--order 
perturbation theory are related to $B$ RPA matrix elements. The RPA residual interaction (second derivative of the Hartree-Fock energy 
functional \cite{rs}, which automatically includes the rearrangement terms) has thus 
to be used to calculate the second--order 
energy correction. As a first approximation, as done in Refs. \cite{moghra2012-1} and \cite{moghra2012-2}, we neglect rearrangement terms. We then include them (together with the effective mass) in the last part of the article for symmetric and neutron matter and provide sets of parameters where their effect is taken into account. 
Such sets of parameters may be considered a very reasonable starting point to construct a bridge between infinite matter, where our effective regularized interactions are presently adjusted, and finite nuclei, where we eventually plan to employ them. 

We analyze here the second--order integrals using two choices for the
momentum cutoff. While the numerical Monte Carlo integration is performed
with a cutoff ($\Lambda$) on the transferred momentum $\mathbf{q}$, it turns
out that the use of a cutoff ($\lambda$) on the outgoing relative momentum $%
\mathbf{k^{\prime}}$ is more convenient for the analytical derivation. This
choice is analogous to that adopted for instance in the low--momentum
interaction $V_{low-k}$ \cite{bogner}. The analytical derivation is
performed here only for symmetric and pure neutron matter. For asymmetric
matter, we solve numerically the second--order integrals with the Monte
Carlo method. To present a coherent analysis of the obtained results and
discuss figures with a unique choice for the momentum cutoff, we show in
this article, in the case of the cutoff regularization, second--order EOS's
obtained in all cases numerically, with a cutoff equal to $\Lambda$.

The article is organized as follows. In Sect. II the cutoff regularization
is discussed. First, the analytical expressions of the second--order EOS's
are shown in the cases of symmetric and pure neutron matter (Sect. II.A).
Numerical results for the second--order EOS's, obtained with the Monte Carlo
method, are then shown for several values of the cutoff $\Lambda$ for
symmetric, neutron, and asymmetric matter. Adjustments of parameters are
presented and discussed (Sect. II.B). In Sect. III the
dimensional--regularized results are illustrated and the adjustment of the
parameters is discussed also in this case. 
Section IV illustrates results obtained for symmetric and neutron matter in the case where  
the approximation $m^*=m$ is not employed ($m^*$ is taken equal to its mean--field value) and  rearrangement terms are included. Sets of parameters are provided. 
We draw conclusions in Sect. V.
Appendix A lists the factors appearing in front of the different types of
integrals which are solved by the Monte Carlo method.

\section{Cutoff regularization}

\subsection{Analytical derivation of the second--order contribution for
symmetric and neutron matter}

We start by writing the standard Skyrme interaction, 
\begin{widetext}
\begin{equation}
v(\mathbf{k},\mathbf{k^{\prime
}}) =t_{0}(1+x_{0}P_{\sigma })+\frac{1}{2}t_{1}(1+x_{1}P_{\sigma })(\mathbf{k%
}^{\prime 2}+\mathbf{k}^{2}) +t_{2}(1+x_{2}P_{\sigma })\mathbf{k}^{\prime
}\cdot \mathbf{k}   
+\frac{1}{6}t_{3}(1+x_{3}P_{\sigma })\rho ^{\alpha }, 
\label{sk}
\end{equation}
\end{widetext}where we adopt the following convention, 
\begin{equation}
v(\mathbf{k},\mathbf{k^{\prime }})=\int \int d^{3}\mathbf{r}d^{3}\mathbf{%
r^{\prime }}e^{-i\mathbf{k}\cdot \mathbf{r}}v(\mathbf{r},\mathbf{r^{\prime }}%
)e^{i\mathbf{k^{\prime }}\cdot \mathbf{r^{\prime }}}.  \label{transfo}
\end{equation}%
As already anticipated, we have omitted the spin--orbit and tensor parts,
for simplicity. The parameters ($t_{i}$, $x_{i}$, and $\alpha $) are in this
case nine. $P_{\sigma }$ is the spin--exchange operator, $P_{\sigma }=\frac{1%
}{2}(1+\sigma _{1}\cdot \sigma _{2})$. The mean--field or first--order
contribution to the EOS of symmetric matter is the well--known expression 
\begin{equation*}
\frac{E_{sym}}{A}^{(1)}=\frac{3}{10}\frac{\hbar ^{2}}{m}\left( \frac{3\pi
^{2}}{2}\right) ^{\frac{2}{3}}\rho ^{\frac{2}{3}}+\frac{3}{8}t_{0}\rho +%
\frac{3}{80}\left( \frac{3\pi ^{2}}{2}\right) ^{\frac{2}{3}}\Theta _{S}\rho
^{\frac{5}{3}}+\frac{1}{16}t_{3}\rho ^{\alpha +1},
\end{equation*}%
where $\Theta _{S}=3t_{1}+t_{2}(5+4x_{2})$. For neutron matter, the
first--order contribution to the EOS is written as 
\begin{equation*}
\frac{E_{neutr}}{A}^{(1)}=\frac{3}{10}\frac{\hbar ^{2}}{m}\left( 3\pi
^{2}\right) ^{\frac{2}{3}}\rho ^{\frac{2}{3}}+\frac{1}{4}t_{0}\rho (1-x_{0})+%
\frac{3}{40}\left( 3\pi ^{2}\right) ^{\frac{2}{3}}(\Theta _{S}-\Theta
_{V})\rho ^{\frac{5}{3}}+\frac{1}{24}t_{3}\rho ^{\alpha +1}(1-x_{3}),
\end{equation*}%
with $\Theta _{V}=t_{1}(2+x_{1})+t_{2}(2+x_{2})$. In symmetric matter, the
total density and the Fermi momentum $k_{F}$ are related by the relation $%
k_{F}=\left( \frac{3\pi ^{2}}{2}\rho \right) ^{1/3}$. Neutron and proton
Fermi momenta are in this case equal to $k_{F}$. In neutron matter, the
Fermi momentum $k_{F_{N}}$ is related to the total density (which is equal
to the neutron density) by the relation $k_{F_{N}}=\left( 3\pi ^{2}\rho
\right) ^{1/3}$.

The second--order contribution is calculated by solving the integral 
\begin{equation}
\Delta E^{(2)} = -\frac{1}{4} \frac{\Omega^3}{(2\pi)^9} \int d^3\mathbf{k_1}
\int d^3\mathbf{k_2} \int d^3\mathbf{q} \frac{|<\mathbf{k_1} \mathbf{k_2}
|V| \mathbf{k_1^{\prime}} \mathbf{k_2^{\prime}>}|^2} {\epsilon _{1}^{\prime
}+\epsilon _{2}^{\prime }-\epsilon _{1}-\epsilon _{2}},  \label{inte1}
\end{equation}%
where $V=v/\Omega$, $\Omega$ is the box volume where the wave functions are
normalized, $v$ is the Skyrme interaction written in Eq. (\ref{sk}), $%
\mathbf{k_{1}^{\prime }} =\mathbf{q}+\mathbf{k_{1}},\;\mathbf{k_{2}^{\prime }%
}=\mathbf{k_{2}}-\mathbf{q}$, and $\mathbf{q}$ is the transferred momentum.
Eq. (\ref{inte1}) may be written as 
\begin{equation}
\Delta E^{(2)} = \frac{1}{4} \frac{\Omega}{(2\pi)^9} \int d^3\mathbf{k_1}
\int d^3\mathbf{k_2} \int d^3\mathbf{q} |<\mathbf{k_1} \mathbf{k_2}
|v| \mathbf{k_1^{\prime}} \mathbf{k_2^{\prime}}>|^2 G,  \label{integral}
\end{equation}%
where we have introduced the propagator $G$, 
\begin{eqnarray}
G &=&\frac{-1}{\epsilon _{1}^{\prime }+\epsilon _{2}^{\prime }-\epsilon
_{1}-\epsilon _{2}},\;\;\epsilon _{i}^{(^{\prime })}=\frac{\hbar
^{2}k_{i}^{(^{\prime })2}}{2m_{i}^{\ast }},  \label{prop}
\end{eqnarray}%
with 
\begin{eqnarray}
\mid \mathbf{k_{1}}\mid &<&k_{F1},\;\;\mid \mathbf{k_{2}}\mid <k_{F2}, 
\notag \\
\mid \mathbf{q+k_{1}}\mid &>&k_{F1},\;\;\mid \mathbf{k_{2}-q}\mid >k_{F2}.
\end{eqnarray}%
In Eq. (\ref{prop}), $m_{i}^{\ast }$ represents the nucleonic effective
mass. In this work, we first take  the approximation $m_{i}^{\ast}=m$  
for simplicity. Such approximation will not be adopted in Sect. IV. In the integral of Eq. (\ref{integral}), $\mathbf{k_1}$ and $\mathbf{%
k_2}$ lie inside the Fermi spheres associated to $k_{F1}$ and $k_{F2}$,
respectively, and the integrals on $\mathbf{k_1}$ and $\mathbf{k_2}$ do not
diverge. An ultraviolet divergence appears in the computation of the
integral on the transferred momentum $\mathbf{q}$ and a cutoff $\Lambda$ is
put in such integral as a regulator.

We now introduce the incoming $\mathbf{k}$ and outgoing $\mathbf{k^{\prime}}$
relative momenta, appearing in the Skyrme interaction, Eq. (\ref{sk}), and
related to $\mathbf{k_{1}}$, $\mathbf{k_{2}}$, and $\mathbf{q}$ by 
\begin{equation}
\mathbf{k}=\frac{\mathbf{k_1}-\mathbf{k_2}}{2}, \; \mathbf{k^{\prime}}=\frac{%
\mathbf{k_1^{\prime}}-\mathbf{k_2^{\prime}}}{2}= \frac{\mathbf{k_1}-\mathbf{%
k_2}}{2}+\mathbf{q}.  \label{varia}
\end{equation}
In this work, the analytical derivation of the second--order contribution to
the EOS has been done following Ref. \cite{baranger}. This specific
derivation can be adapted only to the cases of symmetric and pure neutron
matter, where there is a unique Fermi momentum $k_{F1}= k_{F2}$, and not
to the case of asymmetric matter, where $k_{F1} \ne k_{F2}$. For asymmetric
matter, the EOS is computed numerically by a Monte Carlo integration and
discussed in Sect. II.B together with all the other numerical results.
In the present section, as well as in the following, we neglect the rearrangement terms associated to the 
density--dependent part of the interaction. Such terms will be included in Sect. IV.

\subsubsection{Symmetric matter}

In symmetric matter as well as in neutron matter (next subsection) it is advantageous to
perform the change of variables given by Eq. (\ref{varia}) to write the
propagator as%
\begin{equation}
G=\frac{-m}{\hbar^2(k^{\prime }{}^{2}-k^{2})}.
\end{equation}%
Starting from Eq. (\ref{integral}), dividing by the number of particles $%
A=\Omega\rho$, writing explicitly the sums over spin and isospin and the
direct and exchange terms, the second--order correction to the EOS is equal
to 
\begin{eqnarray}
\frac{E_{sym}^{(2)}}{A} &=& -\frac{3m}{32 (2\pi)^7 \hbar^2 k_{F}^{3}}%
\sum\limits_{STM_{S}M_{S^{\prime }}}(2T+1)\int \int \int d^{3}\mathbf{K}d^{3}%
\mathbf{k}d^{3}\mathbf{k}^{\prime }  \notag \\
&&\times \frac{|\langle X_{M_{S}}^{S}|v_{ST}(\mathbf{k},\mathbf{k}^{\prime
})-(-)^{S+T}v_{ST}(\mathbf{k},-\mathbf{k}^{\prime })|X_{M_{S^{\prime
}}}^{S}\rangle |^{2}}{(k^{\prime 2}-k^{2})}.  \label{eos1}
\end{eqnarray}
Here $S$ and $T$ are the total spin and isospin, respectively, $%
M_{S^{(\prime )}}$ is the projection of $S$ on the $z$--axis, and $%
X_{M_{S^{(\prime) }}}^{S}$ is the two--body spin state. The interaction $%
v_{ST}$ is always the interaction $v$ of Eq. (\ref{sk}), after having 
evaluated the expectation value in the isospin state, and where we have
explicitly indicated spin and isospin labels for convenience. Note that the
additional factor $1/(2\pi)^6$ in Eq. (\ref{eos1}), with respect to the
corresponding expression in Ref. \cite{baranger}, comes from the different
convention adopted in Eq. (\ref{transfo}). The two terms in Eq. (\ref{eos1})
represent the direct and exchange contributions. We have introduced a third
variable $\mathbf{K}$, which is chosen in the same way as in Ref. \cite%
{baranger}, that is $\mathbf{K}\equiv \mathbf{k}_{1}+\mathbf{k}_{2}$. Since
both $\mathbf{k_1}$ and $\mathbf{k_2}$ lie inside the Fermi sphere
associated to $k_F$, the integrals on the incoming relative momentum $%
\mathbf{k}$ and on $\mathbf{K}$ do not diverge. A regulator must however be
put on the diverging integral in $\mathbf{k^{\prime}}$ (cutoff $\lambda$).
If this cutoff is chosen smaller than $k_F$, also the integral on the
incoming momentum must be regulated by the same cutoff $\lambda$.

The interaction can be expanded in partial waves,%
\begin{equation}
v_{ST}(\mathbf{k},\mathbf{k}^{\prime })=\sum\limits_{JM_{J}ll^{\prime
}}v_{ST,ll^{\prime }}^{J}(k,k^{\prime })y_{lS}^{JM_{J}}(\widehat{\mathbf{k}}%
)[y_{l^{\prime }S}^{JM_{J}}(\widehat{\mathbf{k}}^{\prime })]^{\dag },
\label{partial}
\end{equation}
where $y_{lS}^{JM_{J}}$ is written as 
\begin{equation}
y_{lS}^{JM_{J}}(\widehat{\mathbf{k}})=\sum\limits_{m_{l},M_{S}}\langle
JM_{J}|lSm_{l}M_{S}\rangle Y_{lm_{l}}(\widehat{\mathbf{k}})X_{M_{S}}^{S}.
\end{equation}%
In general, $l$ and $l^{\prime}$ in Eq. (\ref{partial}) must have the same
parity. By imposing antisymmetrization, it holds: 
\begin{equation}
(-1)^{l+S+T}=(-1)^{l^{\prime}+S+T}=-1.  \notag
\end{equation}
This implies that the exchange term in Eq. (\ref{eos1}) is equal to the
direct term. After some manipulations, and by evaluating the spin matrix
elements, Eq. (\ref{eos1}) becomes 
\begin{eqnarray}
\frac{E_{sym}^{(2)}}{A} &=&-\frac{3m}{16 (2\pi)^8 \hbar^2 k_{F}^{3}}
\sum_{STJ\bar{J}LMll^{\prime}\bar{l}\bar{l^{\prime}}} \delta_{STl} \delta_{ST%
\bar{l}} (2T+1) (2J+1) (2\bar{J}+1) (2L+1)^{-1}  \notag \\
&\times& \left[ (2l+1) (2l^{\prime}+1) (2\bar{l}+1)(2\bar{l^{\prime}}+1)%
\right]^{1/2} < L0|l\bar{l}00> < L0|l^{\prime}\bar{l^{\prime}}00> W(J\bar{J}l%
\bar{l};LS) W(J\bar{J}l^{\prime}\bar{l^{\prime}};LS)  \notag \\
&\times& \int \int \int d^{3}\mathbf{K}d^{3}\mathbf{k}d^{3}\mathbf{k}%
^{\prime} Y_{LM}(\hat{k}) Y_{LM}^*(\hat{k^{\prime}}) v_{ST,ll^{\prime
}}^{J}(k,k^{\prime }) v_{ST,\bar{l}\bar{l^{\prime }}}^{\bar{J}}(k,k^{\prime
}) (k^{\prime 2}-k^2)^{-1},  \label{partialw}
\end{eqnarray}
with 
\begin{equation}
\delta_{STl}=\frac{1}{2}\left[1-(-1)^{S+T+l}\right].
\end{equation}
In Eq. (\ref{partialw}), $W$ indicates Racah coefficients. In our case, the
interaction is diagonal in $l$ and independent on $J$, that is, $%
v_{ST,ll^{\prime }}^{J}(k,k^{\prime })=\delta _{ll^{\prime
}}v_{S,T,l}(k,k^{\prime })$. The product $\delta_{STl} \delta_{ST\bar{l}}$
in Eq. (\ref{partialw}) implies that $l$ and $\bar{l}$ must have the same
parity. This means that, for a given $(S,T)$, only even--even and odd--odd
partial waves of the interaction may mix at second order. The Skyrme
interaction of Eq. (\ref{sk}) contains only one type of even waves, the $s$%
--wave $t_{0}$, $t_{3}$, and $t_{1}$ terms, and one type of odd waves, the $%
p $--wave $t_{2}$ term. Consequently, $l$ and $\bar{l}$ must be the same
(equal to 0 or 1) and the only quadratic terms that enter in the
second--order contribution are proportional to $t_{0}^{2}$, $t_{3}^{2}$, $%
t_{1}^{2}$, $t_{0}t_{3}$, $t_{0}t_{1}$, $t_{3}t_{1} $, and $t_{2}^{2}$ (the
only possible values for $L$ are $L=$ 0 and 2 for the $p$--wave case and $%
L=0 $ for the $s$-wave case). Interference terms proportional to $t_{0}t_{2}$%
, $t_{3}t_{2}$, and $t_{1}t_{2}$ are absent in the EOS of symmetric and pure
neutron matter. We stress that, on the other side, such
interference terms are present in the EOS of asymmetric matter. This occurs
due to the different Fermi momenta between neutrons and protons in
asymmetric matter. Also the different effective masses of neutrons and
protons would be responsible for such interference terms.

We write now explicitly the squares of the interaction $v^2_{S,T,l}(k,k^{%
\prime})$ in the different channels. For the isovector case $T=1$, one has $%
S=0$ ($P_{\sigma }=-1$) for $l=0$, and $S=1$ ($P_{\sigma }=1$) for $l=1$.
The square of the interaction is then written as

\begin{widetext}
\begin{eqnarray}
\nonumber
v_{S=0,T=1,l=0}^2 (k,k') &=&(4\pi)^2 [t_{0}^{2}(1-x_{0})^{2}+\frac{1}{4}%
t_{1}^{2}(1-x_{1})^{2}(k^{\prime 4}+2k^{\prime 2}k^{2}+k^{4})+\frac{1}{36}%
t_{3}^{2}(1-x_{3})^{2}\rho ^{2\alpha }  \\
\nonumber
&&+t_{0}(1-x_{0})t_{1}(1-x_{1})(k^{\prime 2}+k^{2})+\frac{1}{6}t_{1}
(1-x_{1})(k^{\prime
2}+k^{2})t_{3}(1-x_{3})\rho ^{\alpha }  \notag \\
&&+\frac{1}{3}t_{0}(1-x_{0})t_{3}(1-x_{3})\rho ^{\alpha }],  \label{vsquare1}
\end{eqnarray}%
and 
\begin{equation}
v_{S=1,T=1,l=1}^{2}(k,k') =\frac{(4\pi)^2}{9}t_{2}^{2}(1+x_{2})^{2}(k^{\prime }k)^{2},   \label{vsquare2}
\end{equation}%
\end{widetext}for the two cases, respectively. Note that the factors $(4\pi)^2$
and $(4\pi)^2/9$ in Eqs. (\ref{vsquare1}) and (\ref{vsquare2}), respectively, come from the partial wave expansion, Eq. (\ref{partial}), of the
Skyrme interaction. In the isoscalar case $T=0$, one has $S=1$ ($P_{\sigma
}=1$) for $l=0$ and $S=0$ ($P_{\sigma }=-1$) for $l=1$. The expressions for
the square of the interaction $v_{S=1,T=0,l=0}^{2}(k,k^{\prime })$ and $%
v_{S=0,T=0,l=1}^{2}(k,k^{\prime })$ may be obtained from Eqs. (\ref{vsquare1}%
) and (\ref{vsquare2}), respectively, by substituting $(1+x_{i})$ to $%
(1-x_{i})$ in Eq. (\ref{vsquare1}) and $(1-x_{2})$ to $(1+x_{2})$ in Eq. (%
\ref{vsquare2}).

We perform a change of variables to use dimensionless vectors, 
\begin{equation}
\mathbf{y}=\frac{\mathbf{k}}{k_{F}},\;\mathbf{y^{\prime }}=\frac{\mathbf{%
k^{\prime }}}{k_{F}},\;\mathbf{x}=\frac{\mathbf{K}}{2k_{F}}.
\end{equation}%
The new variables $\mathbf{y}$ and $\mathbf{y^{\prime }}$ should satisfy the
conditions 
\begin{equation}
|\mathbf{y}|<1,\;|\mathbf{y^{\prime }}|<\lambda .
\end{equation}%
%
%
We integrate over all angles by using the function, 
\begin{equation}
J_{LM}(x,y,y^{\prime })=\int \int \int d\hat{x}d\hat{y}d\hat{y^{\prime }}%
Y_{LM}(\hat{y})Y_{LM}^{\ast }(\hat{y^{\prime }}).
\end{equation}%
After some manipulations, one can write 
\begin{equation}
J_{LM}(x,y,y^{\prime })=16\pi ^{2}\delta _{M,0}(2L+1)A_{L}(y,x)A_{L}^{\prime
}(y^{\prime },x),
\end{equation}%
where the explicit expressions of the functions $A_{L}^{(^{\prime })}$ are
given by Eqs. (3.16a)-(3.17b) of Ref. \cite{baranger}. The radial
integration on $x$ is done and the functions $I^{(L)}$ are introduced, 
\begin{equation}
I^{(L)}(y,y^{\prime })=\int_{0}^{1}x^{2}dxA_{L}(y,x)A_{L}^{\prime
}(y^{\prime },x).
\end{equation}%
The following expressions for the $s$-- and $p$--wave contributions to the
EOS may be finally written, 
\begin{widetext} 
\begin{eqnarray}
\frac{\Delta E_{sym (l=1)}^{(2)}}{A} &=&-\frac{18mk_{F}^{4}}{4\pi^4 \hbar^2}\sum_{ST}(2T+1)(2S+1)\delta _{ST1}\int \int
dydy^{\prime }\frac{y^{2}y^{\prime 2}}{y^{\prime 2}-y^{2}}\left[
I^{(0)}(y,y^{\prime })+2I^{(2)}(y,y^{\prime })\right]
[v_{S,T,1}(k_{F}y,k_{F}y^{\prime })]^{2},  \notag \\
\frac{\Delta E_{sym (l=0)}^{(2)}}{A} &=&-\frac{18mk_{F}^{4}}{4\pi^4 \hbar^2}\sum_{ST}\delta _{ST0}\int \int dydy^{\prime }%
\frac{y^{2}y^{\prime 2}}{y^{\prime 2}-y^{2}}I^{(0)}(y,y^{\prime
})[v_{S,T,0}(k_{F}y,k_{F}y^{\prime })]^{2}.  \label{a2}
\end{eqnarray}%
\end{widetext}The expressions of the EOS's may be then obtained
analytically. The two terms that should be summed up are 
\begin{widetext}
\begin{equation}
\frac{\Delta E_{sym (l=0)}^{(2)}}{A}=-\frac{mk_{F}^{4}}{110880\hbar^2\pi ^{4}}\left\{ 
\begin{array}{c}
\left[ 
\begin{array}{c}
-6534+1188 ln[2]+3564\lambda -19602\lambda ^{3}-5940\lambda ^{5} \\ 
+(1782-20790\lambda ^{4}) ln[\frac{\lambda -1}{\lambda +1}] \\ 
+(24948\lambda ^{5}-5940\lambda ^{7}) ln[\frac{\lambda ^{2}-1}{\lambda ^{2}}]%
\end{array}%
\right] \widetilde{T}_{03}^{2} \\ 
+\left[ 
\begin{array}{c}
-\newline
14696+2112 ln[2]+5280\lambda -2860\lambda ^{3} \\ 
-\newline
48840\lambda ^{5}-18480\lambda ^{7}+(2640-55440\lambda ^{6}) ln[\frac{%
\lambda -1}{\lambda +1}] \\ 
+(71280\lambda ^{7}-18480\lambda ^{9}) ln[\frac{\lambda ^{2}-1}{\lambda ^{2}}%
]%
\end{array}%
\right] k_{F}^{2}\widetilde{T}_{03}\widetilde{T}_{1} \\ 
+\left[ 
\begin{array}{c}
-9886+1128 ln[2]+\newline
2520\lambda +147\lambda ^{3}-3654\lambda ^{5} \\ 
-\newline
35280\lambda ^{7}-15120\lambda ^{9}+(1260-41580\lambda ^{8}) ln[\frac{%
\lambda -1}{\lambda +1}] \\ 
+(55440\lambda ^{9}-15120\lambda ^{11}) ln[\frac{\lambda ^{2}-1}{\lambda ^{2}%
}]%
\end{array}%
\right] k_{F}^{4}\widetilde{T}_{1}^{2}%
\end{array}%
\right\} 
\label{symml0}
\end{equation}%
and
\begin{equation}
\frac{\Delta E_{sym (l=1)}^{(2)}}{A}=-\frac{mk_{F}^{8}}{73920\hbar^2\pi ^{4}}\left\{ \left[ 
\begin{array}{c}
-1033+156 ln[2]+420\lambda +140\lambda ^{3}-840\lambda ^{5} \\ 
-5880\lambda ^{7}-2520\lambda ^{9}+(-210+6930\lambda ^{8}) ln[\frac{\lambda
-1}{\lambda +1}] \\ 
+(9240\lambda ^{9}-2520\lambda ^{11}) ln[\frac{\lambda ^{2}-1}{\lambda ^{2}}]%
\end{array}%
\right] \widetilde{T}_{2}^{2}\right\},
\label{symml1} 
\end{equation}
where we have introduced the combinations of parameters  
\begin{eqnarray}
\nonumber
\widetilde{T}_{03}^{2} &=&\left[ t_{0}(1-x_{0})+\frac{1}{6}%
t_{3}(1-x_{3})\rho ^{\alpha }\right] ^{2}+\left[ t_{0}(1+x_{0})+\frac{1}{6}%
t_{3}(1+x_{3})\rho ^{\alpha }\right] ^{2}   \\
\nonumber
\widetilde{T}_{1}^{2} &=&\frac{1}{4}t_{1}^{2}\left[
(1-x_{1})^{2}+(1+x_{1})^{2}\right] =\frac{1}{2}t_{1}^{2}(1+x_{1}^{2})
 \\
\nonumber
\widetilde{T}_{03}\widetilde{T}_{1} &=&\frac{t_{1}}{2}\left[ [t_{0}(1-x_{0})+%
\frac{1}{6}t_{3}(1-x_{3})\rho ^{\alpha }](1-x_{1})+[t_{0}(1+x_{0})+\frac{1}{6%
}t_{3}(1+x_{3})\rho ^{\alpha }](1+x_{1})\right]   \\
\nonumber
\widetilde{T}_{2}^{2} &=&[t_{2}^{2}(1-x_{2})^{2}+9t_{2}^{2}(1+x_{2})^{2}]/9 
\\
&=&\frac{2}{9}t_{2}^{2}(5+8x_{2}+5x_{2}^{2}).
\label{tnomsymm}  
\end{eqnarray}%
The asymptotic behavior can be written as a
polynomial form in $\lambda$. One has to sum up the
two terms
\begin{equation}
\frac{\Delta E_{sym (l=0),asympt.}^{(2)}}{A}=-\frac{9mk_{F}^{4}}{2\hbar^2\pi ^{4}}%
\begin{array}{c}
\left[ 
\begin{array}{c}
\frac{k_{F}^{4}\widetilde{T}_{1}^{2}}{360}\lambda ^{5}+(\frac{k_{F}^{2}%
\widetilde{T}_{03}\widetilde{T}_{1}}{108}+\frac{k_{F}^{4}\widetilde{T}%
_{1}^{2}}{240})\lambda ^{3}+(\frac{\widetilde{T}_{03}^{2}}{72}+\frac{%
k_{F}^{2}\widetilde{T}_{03}\widetilde{T}_{1}}{60}+\frac{k_{F}^{4}\widetilde{T%
}_{1}^{2}}{140})\lambda  \\ 
+\frac{44k_{F}^{2}\widetilde{T}_{03}\widetilde{T}%
_{1}(-167+24 ln[2])+k_{F}^{4}\widetilde{T}_{1}^{2}(-4943+564 ln[2])+297%
\widetilde{T}_{03}^{2}(-11+2 ln[2])}{249480} \\ 
-(\frac{\widetilde{T}_{03}^{2}}{240}+\frac{k_{F}^{2}\widetilde{T}_{03}%
\widetilde{T}_{1}}{140}+\frac{k_{F}^{4}\widetilde{T}_{1}^{2}}{270})/\lambda
+O(\lambda ^{-2})%
\end{array}%
\right] 
\end{array}%
\end{equation}%
and 
\begin{equation}
\frac{\Delta E_{sym (l=1),asympt.}^{(2)}}{A}=-\frac{9mk_{F}^{8}}{2\hbar^2\pi ^{4}}%
\begin{array}{c}
\left[ 
\begin{array}{c}
\frac{1}{720}\lambda ^{3}+\frac{1}{560}\lambda +(\frac{-1033+156 ln[2]}{%
332640}) \\ 
-(\frac{1}{1080})/\lambda +O(\lambda ^{-2})%
\end{array}%
\right] \widetilde{T}_{2}^{2}%
\end{array}%
.
\end{equation}%

\subsubsection{Neutron matter}

Note that the triple integral is the same for neutron and symmetric matter. 
The factors are not the same (see the nn contribution in Appendix A), leading to a different 
combination of the Skyrme parameters, and 
 $k_F \rightarrow k_{F_N}$.  Here we report the final result. One has to sum up the two terms   
\begin{equation}
\frac{\Delta E_{neutr (l=0)}^{(2)}}{A}=-\frac{mk_{F_N}^{4}}{166320\hbar^2\pi ^{4}}\left\{ 
\begin{array}{c}
\left[ 
\begin{array}{c}
-6534+1188 ln[2]+3564\lambda -19602\lambda ^{3}-5940\lambda ^{5} \\ 
+(1782-20790\lambda ^{4}) ln[\frac{\lambda -1}{\lambda +1}] \\ 
+(24948\lambda ^{5}-5940\lambda ^{7}) ln[\frac{\lambda ^{2}-1}{\lambda ^{2}}]%
\end{array}%
\right] T_{03}^{2} \\ 
+\left[ 
\begin{array}{c}
-\newline
14696+2112 ln[2]+5280\lambda -2860\lambda ^{3} \\ 
-\newline
48840\lambda ^{5}-18480\lambda ^{7}+(2640-55440\lambda ^{6}) ln[\frac{%
\lambda -1}{\lambda +1}] \\ 
+(71280\lambda ^{7}-18480\lambda ^{9}) ln[\frac{\lambda ^{2}-1}{\lambda ^{2}}%
]%
\end{array}%
\right] k_{F_N}^{2}T_{03}T_{1} \\ 
+\left[ 
\begin{array}{c}
-9886+1128 ln[2]+\newline
2520\lambda +147\lambda ^{3}-3654\lambda ^{5} \\ 
-\newline
35280\lambda ^{7}-15120\lambda ^{9}+(1260-41580\lambda ^{8}) ln[\frac{%
\lambda -1}{\lambda +1}] \\ 
+(55440\lambda ^{9}-15120\lambda ^{11}) ln[\frac{\lambda ^{2}-1}{\lambda ^{2}%
}]%
\end{array}%
\right] k_{F_N}^{4}T_{1}^{2}%
\end{array}%
\right\} 
\label{neutl0}
\end{equation}%
and 
\begin{equation}
\frac{\Delta E_{neutr (l=1)}^{(2)}}{A}=-\frac{mk_{F_N}^{8}}{110880\hbar^2\pi ^{4}}\left\{ \left[ 
\begin{array}{c}
-1033+156 ln[2]+420\lambda +140\lambda ^{3}-840\lambda ^{5} \\ 
-5880\lambda ^{7}-2520\lambda ^{9}+(-210+6930\lambda ^{8}) ln[\frac{\lambda
-1}{\lambda +1}] \\ 
+(9240\lambda ^{9}-2520\lambda ^{11}) ln[\frac{\lambda ^{2}-1}{\lambda ^{2}}]%
\end{array}%
\right] T_{2}^{2}\right\} ,
\label{neutl1}
\end{equation}
\end{widetext}where now the combinations of parameters are defined as 
\begin{eqnarray}
T_{03} &=&t_{0}(1-x_{0})+\frac{1}{6}t_{3}(1-x_{3})\rho ^{\alpha },  \notag \\
T_{1} &=&\frac{1}{2}t_{1}(1-x_{1}),  \notag \\
T_{2} &=&t_{2}(1+x_{2}).
\label{tnomneutr}
\end{eqnarray}%
The asymptotic behavior is written as the sum of the two terms 
\begin{equation}
\frac{\Delta E_{neutr(l=0),asympt.}^{(2)}}{A}=-\frac{3mk_{F_{N}}^{4}}{\hbar
^{2}\pi ^{4}}%
\begin{array}{c}
\left[ 
\begin{array}{c}
\frac{k_{F_{N}}^{4}T_{1}^{2}}{360}\lambda ^{5}+(\frac{%
k_{F_{N}}^{2}T_{03}T_{1}}{108}+\frac{k_{F_{N}}^{4}T_{1}^{2}}{240})\lambda
^{3}+(\frac{T_{03}^{2}}{72}+\frac{k_{F_{N}}^{2}T_{03}T_{1}}{60}+\frac{%
k_{F_{N}}^{4}T_{1}^{2}}{140})\lambda \\ 
+\frac{%
44k_{F_{N}}^{2}T_{03}T_{1}(-167+24ln[2])+k_{F_{N}}^{4}T_{1}^{2}(-4943+564ln[2])+297T_{03}^{2}(-11+2ln[2])%
}{249480} \\ 
-(\frac{T_{03}^{2}}{240}+\frac{k_{F_{N}}^{2}T_{03}T_{1}}{140}+\frac{%
k_{F_{N}}^{4}T_{1}^{2}}{270})/\lambda +O(\lambda ^{-2})%
\end{array}%
\right]%
\end{array}%
\end{equation}%
and 
\begin{equation}
\frac{\Delta E_{neutr(l=1),asympt.}^{(2)}}{A}=-\frac{3mk_{F_{N}}^{8}}{\hbar
^{2}\pi ^{4}}%
\begin{array}{c}
\left[ 
\begin{array}{c}
\frac{1}{720}\lambda ^{3}+\frac{1}{560}\lambda +(\frac{-1033+156ln[2]}{332640%
}) \\ 
-(\frac{1}{1080})/\lambda +O(\lambda ^{-2})%
\end{array}%
\right] T_{2}^{2}%
\end{array}%
.
\end{equation}%
One may note that in both EOS's (symmetric and neutron matter) the
divergence is linear in $\lambda $ if only the $t_{0}-t_{3}$ part of the
interaction is taken and goes like $\lambda ^{5}$ if the other terms of the
interaction are also included, as was already indicated in Refs. \cite%
{moghraprl} and \cite{moghra2012-1}. The strongest divergence is provided
by the $t_{1}$ term.

\begin{figure}[tbp]
\includegraphics[scale=0.35]{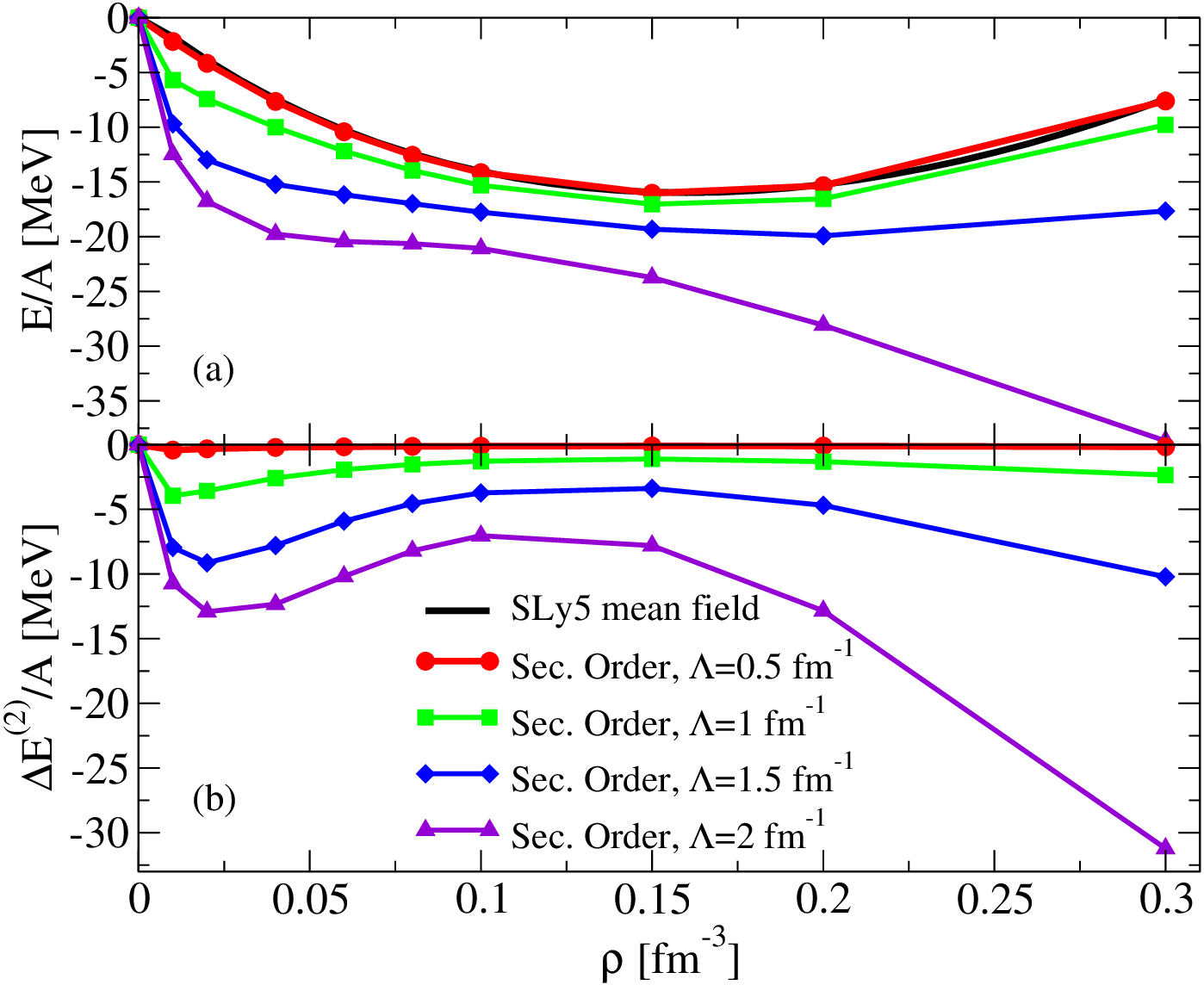}
\caption{(Color online) (a) Second--order EOS of symmetric matter computed
for several values of the cutoff $\Lambda$ and compared with the mean--field
EOS; (b) Second--order correction to the energy per particle for symmetric
matter. The used parameters are those of SLy5.}
\label{two}
\end{figure}

\begin{figure}[tbp]
\includegraphics[scale=0.35]{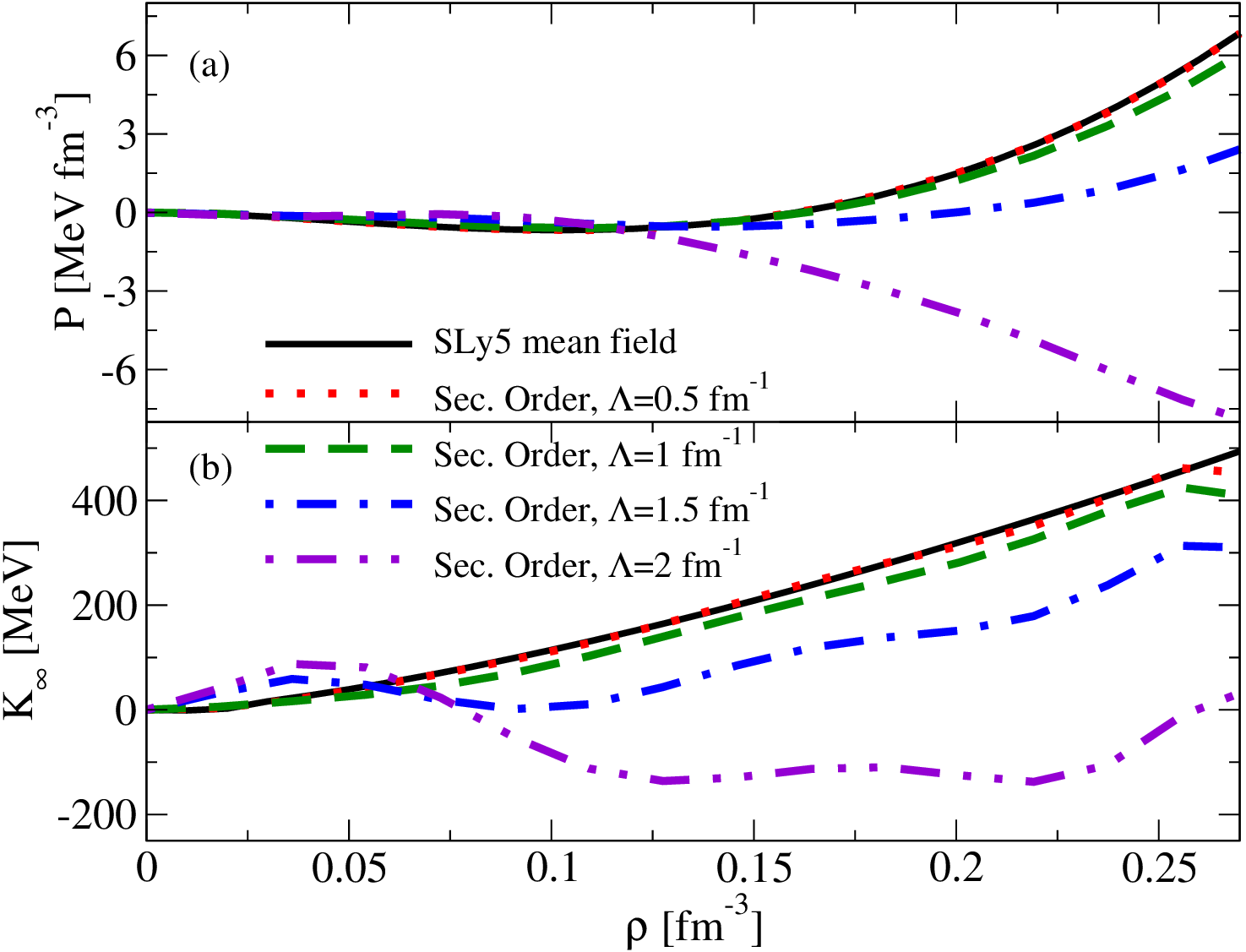}
\caption{(Color online) (a) Second--order pressure; (b) second--order
incompressibility modulus. The used parameters are those of SLy5.}
\label{three}
\end{figure}

\begin{figure}[tbp]
\includegraphics[scale=0.35]{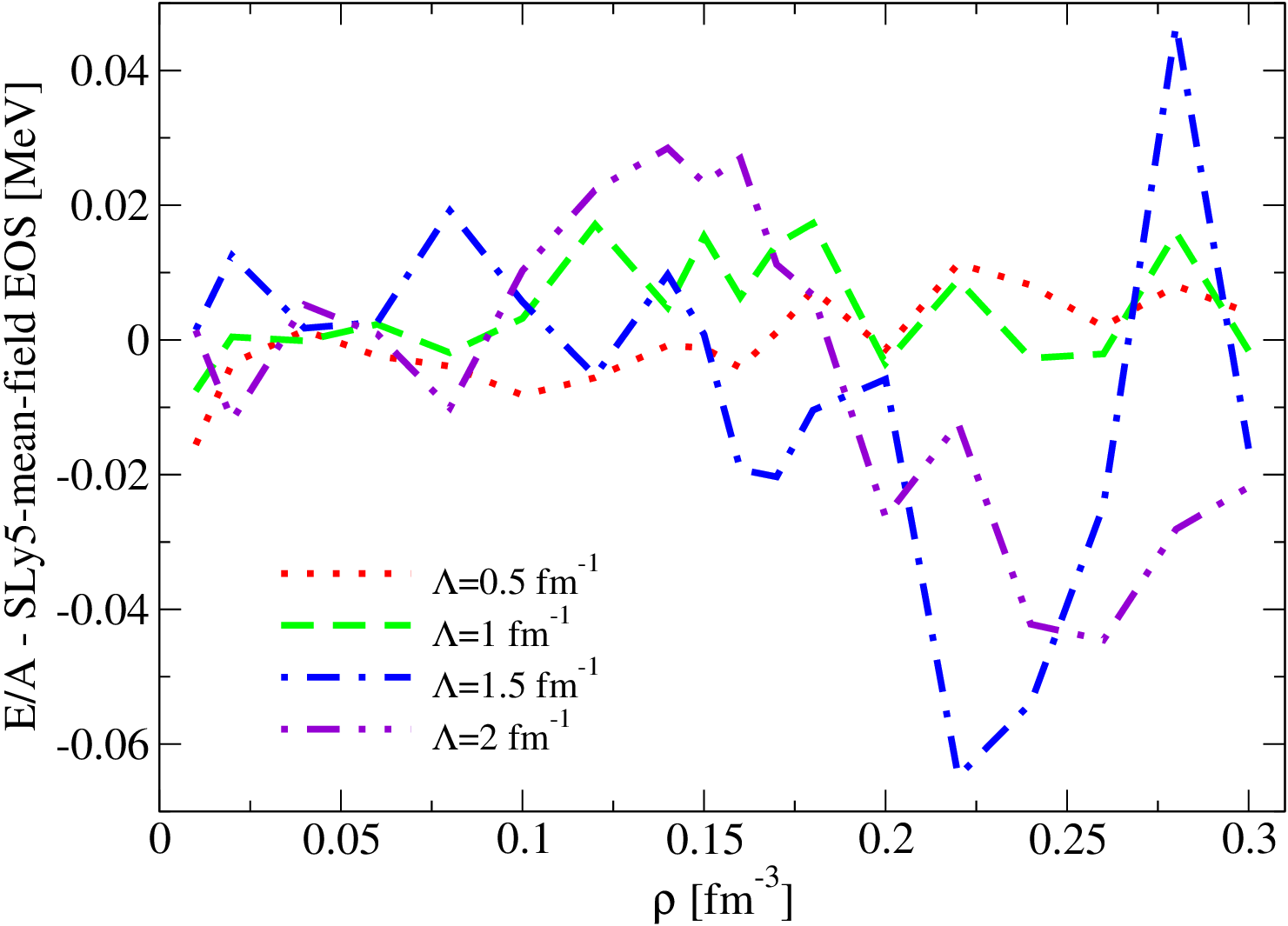}
\caption{(Color online) Difference between the refitted second--order and
the mean--field EOS for symmetric matter.}
\label{four-1}
\end{figure}

\begin{figure}[tbp]
\includegraphics[scale=0.35]{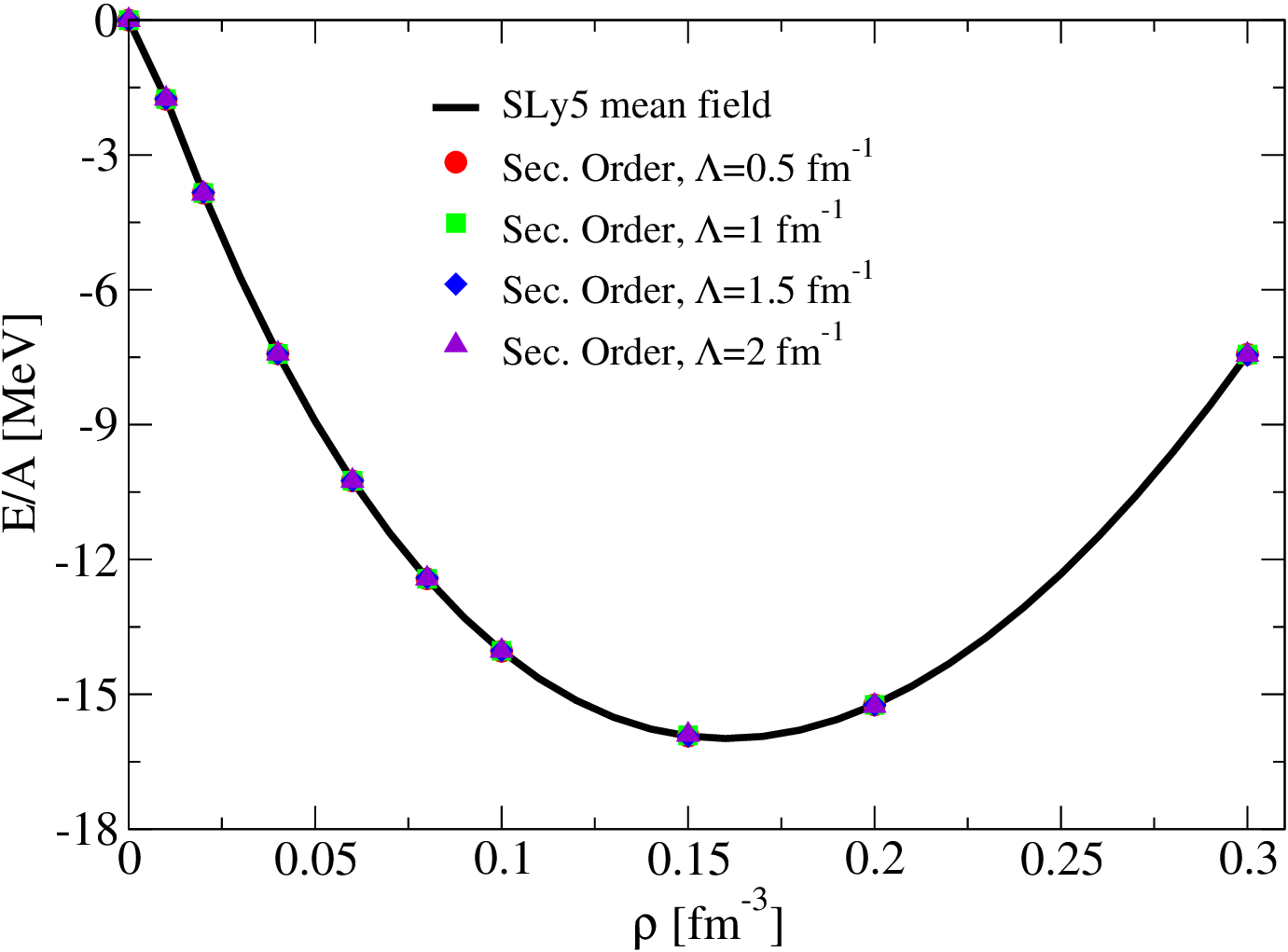}
\caption{(Color online) Second--order refitted EOS compared with the
SLy5--mean--field EOS for symmetric matter.}
\label{four-2}
\end{figure}

\subsection{Numerical results and fits of parameters for symmetric,
asymmetric and pure neutron matter}

We solve the second--order integrals for symmetric, neutron, and asymmetric matter in the
illustrative case $\delta=0.5$, where $\delta$ is the asymmetry parameter $%
\delta=(\rho_n-\rho_p)/(\rho_n+\rho_p)$, and $\rho_n$ and $\rho_p$ are the
neutron and proton densities, respectively. Several types of integrals are
solved numerically (according to the specific second--order contribution).
The factors for which such integrals are multiplied are shown in Appendix A
for all terms.

The adjustment of the parameters is done in all cases on the benchmark 
SLy5--mean--field EOS \cite{sly5}; we use 9 points, 7 of
them located up to 0.16 fm$^{-3}$ and 2 of them located at higher densities,
between 0.16 and 0.3 fm$^{-3}$. The $\chi^2$ values are calculated as $%
\chi^2=1/(N-1)\sum_{i}(E_i-E_{i,ref})^2/\Delta E_i^2$, where $N$ is the
number of points on which the adjustment is done, the sum runs over this
number, $E_{i,ref}$ is the benchmark value corresponding to the point $i$,
and $\Delta E_i$ are all chosen equal to 1\% of the reference value. This
means that, if the $\chi^2$ is less than 1, the average discrepancy between
the adjusted curve and the benchmark EOS is less than 1\% .

\subsubsection{Symmetric matter and incompressibility modulus}

We plot in the upper panel of Fig. \ref{two} the EOS of symmetric matter
calculated up to second order for several cutoff values $\Lambda$. These
curves are compared to the benchmark EOS. All the second--order curves are
obtained by using the same parameters of SLy5. In the lower panel, we plot
only the second--order correction. The ultraviolet divergence is well
visible, especially at densities larger than the saturation density.
Starting from some values of the cutoff between 1.5 and 2 fm$^{-1}$, one
observes that the EOS decreases (instead of increasing) at large densities.

The pressure $P$ and the incompressibility modulus $K$ may be computed from
the EOS as first and second derivatives, respectively, that is, 
\begin{equation}
P(\rho,\Lambda)=\rho^2\frac{d}{d\rho}\frac{E}{A}(\rho,\Lambda)
\end{equation}
and 
\begin{equation}
K(\rho,\Lambda)=9\rho^2\frac{d^2}{d\rho^2}\frac{E}{A}(\rho,\Lambda).
\end{equation}
The second--order pressure (upper panel) and incompressibility modulus
(lower panel) calculated with the parameters of the interaction SLy5 are
displayed in Fig. \ref{three} and compared with the corresponding
mean--field curves.

The adjustment of the nine Skyrme parameters is then performed and Figs. \ref%
{four-1} and \ref{four-2} show the curves obtained with the refitted
parameters. Figure \ref{four-1} presents the difference between the refitted
second--order curve and the SLy5--mean--field curve, whereas the absolute
values are displayed in Fig. \ref{four-2}. The saturation density is, for
all values of the cutoff, the same as the benchmark one, that is 0.16 fm$%
^{-3}$. The refitted parameters are listed in Table I together with the $%
\chi^2$ values. The quality of the fit looks remarkably good as indicated by
the $\chi^2$ values which are, in all cases, not larger than $\sim 3 \times
10^{-2}$.

\begin{figure}[tbp]
\includegraphics[scale=0.35]{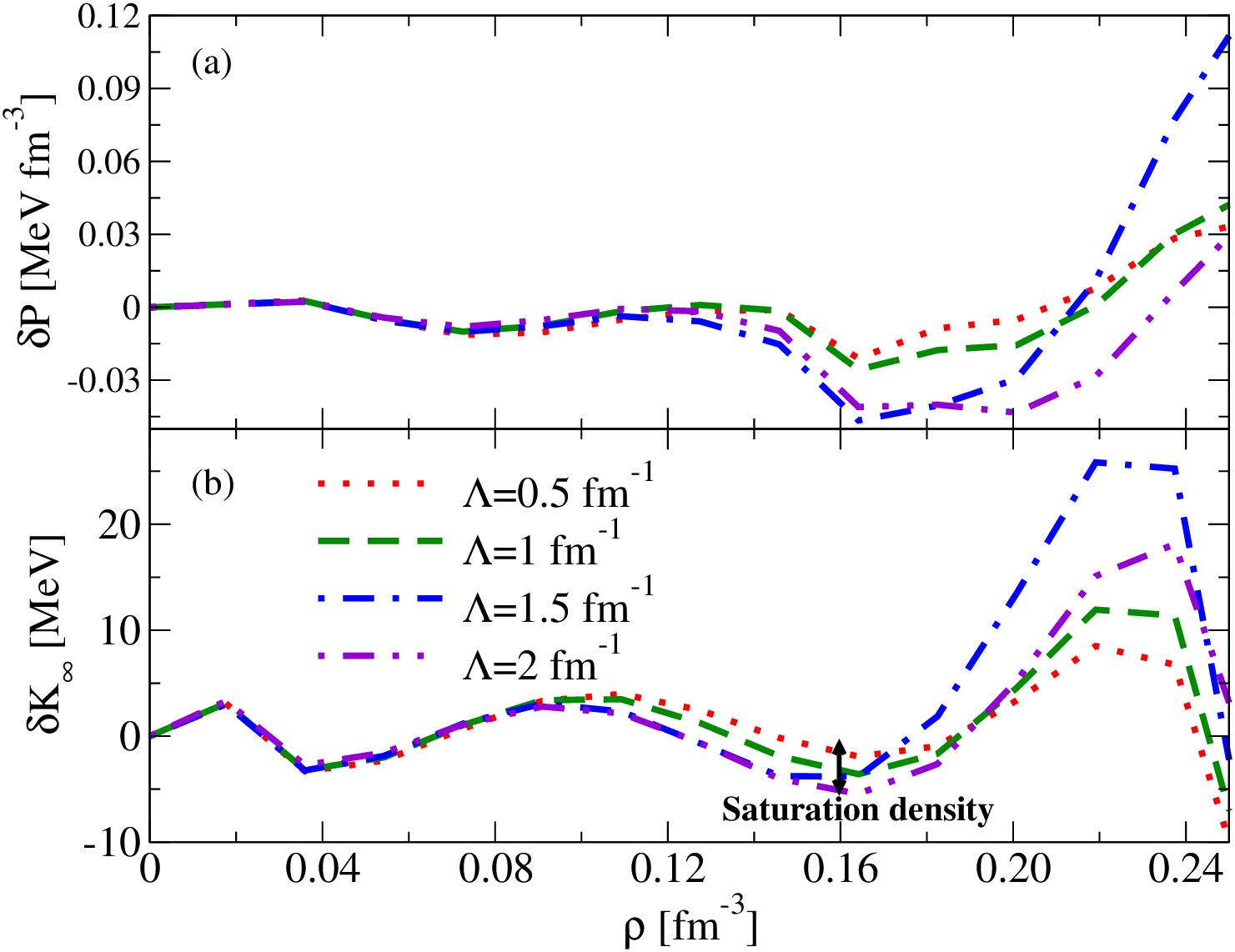}
\caption{(Color online) (a) Difference between the second--order pressure
and the mean--field--SLy5 value; (b) Same as in panel (a), but for the
incompressibility modulus. The black arrow indicates the maximum deviation
of the compressibility from the mean--field--SLy5 value at the saturation
point. }
\label{five-1}
\end{figure}

\begin{figure}[tbp]
\includegraphics[scale=0.35]{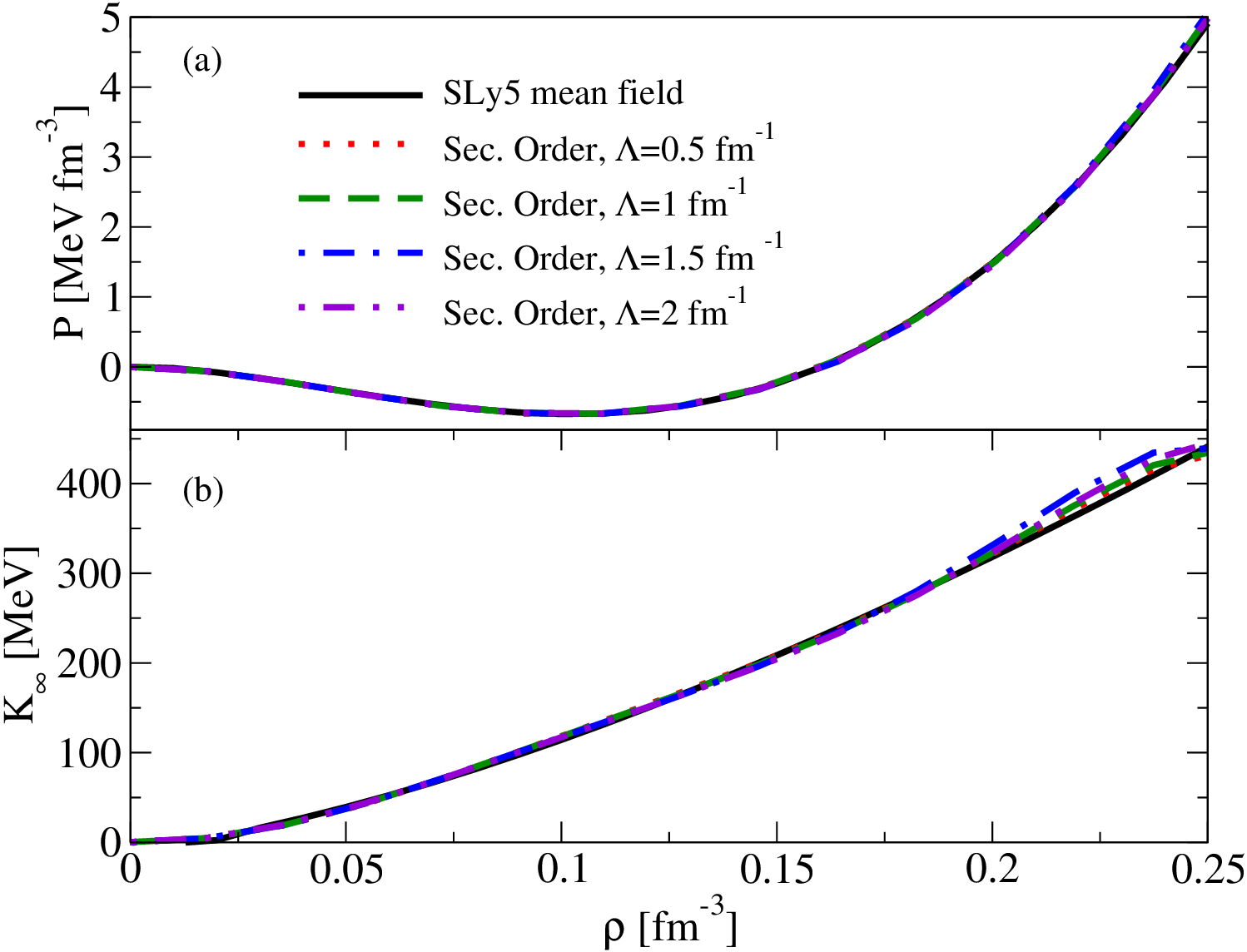}
\caption{(Color online) (a) Second--order pressure compared to the
SLy5-mean--field curve; (b) Same as in panel (a), but for the
incompressibility modulus. }
\label{five-2}
\end{figure}

The pressure and incompressibility modulus are then computed at second
order, this time with the new values for the parameters. Figure \ref{five-1}
displays the difference with respect to the mean--field curves for the
pressure (a) and the incompressibility modulus (b). Figure \ref{five-2}
shows the absolute curves compared with the mean--field ones. We observe
that the maximum deviation of the incompressibility modulus from the
SLy5--mean--field value at the saturation density ($\sim $ 230 MeV) is only $%
\sim $ 5 MeV. This deviation is indicated by a black arrow in the lower
panel of Fig. \ref{five-1}. From the upper panel one can note that the
pressure is non strictly equal to zero at the saturation density. The
deviation in the derivative is however very small and accounts for very
small variations in the saturation density for the different fits.

\subsubsection{Neutron matter}

\begin{figure}[tbp]
\includegraphics[scale=0.35]{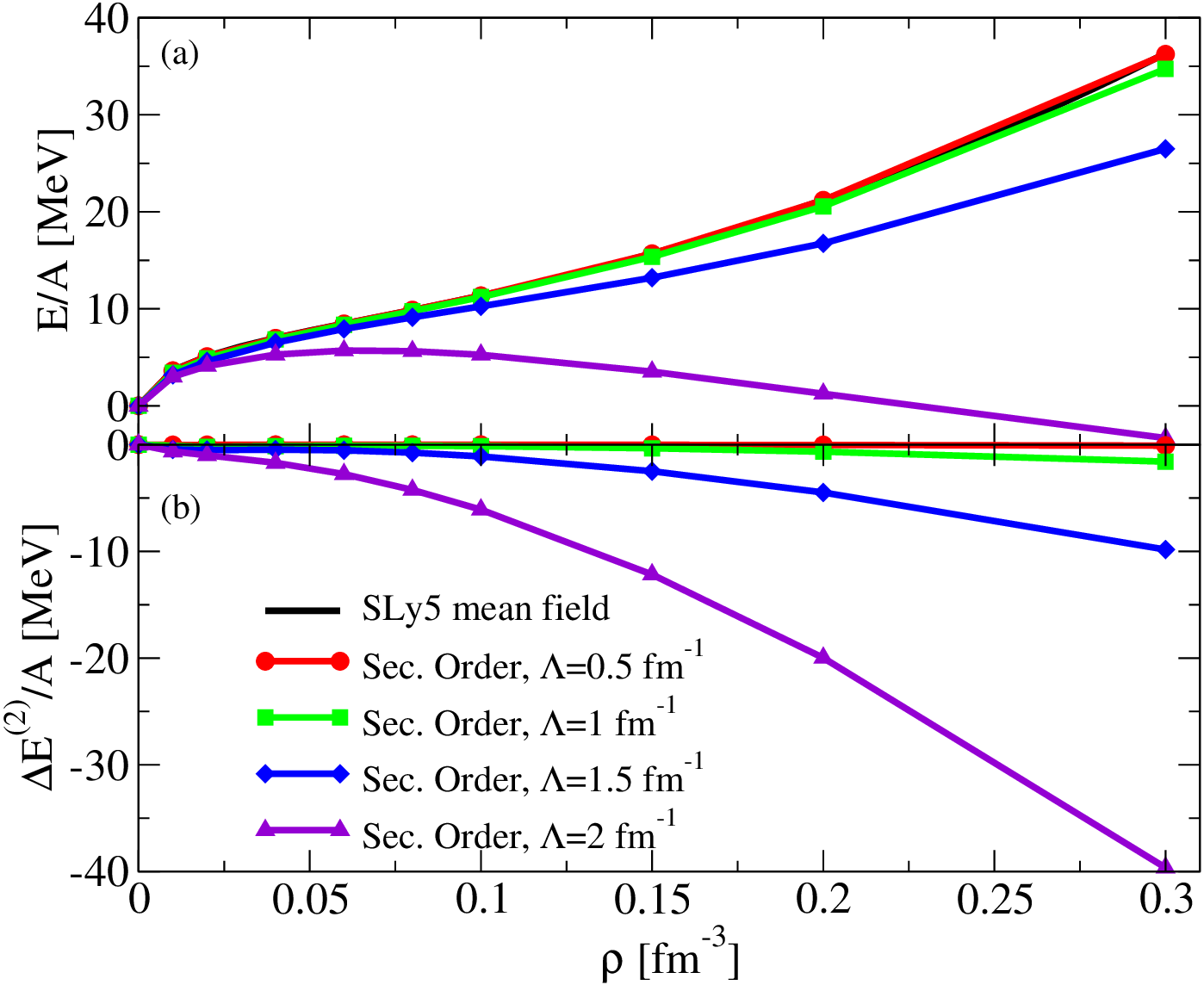}
\caption{(Color online) (a) Second--order EOS of neutron matter for several
values of the cutoff, calculated with the SLy5 parameters, and
SLy5--mean--field EOS; (b) second--order correction.}
\label{six}
\end{figure}

\begin{figure}[tbp]
\includegraphics[scale=0.35]{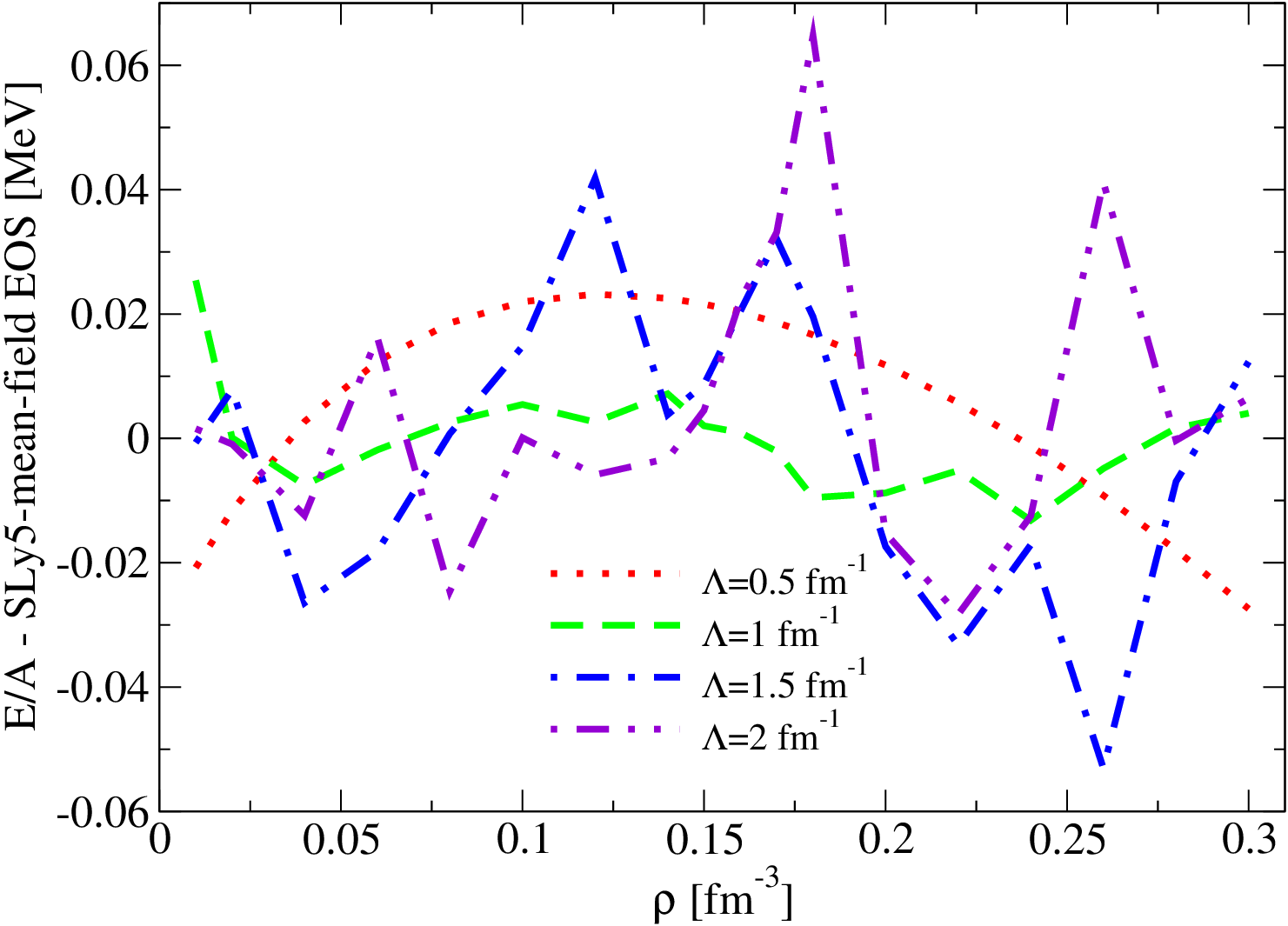}
\caption{(Color online) Difference between the refitted EOS and the
SLy5--mean--field EOS for neutron matter.}
\label{seven-1}
\end{figure}

\begin{figure}[tbp]
\includegraphics[scale=0.35]{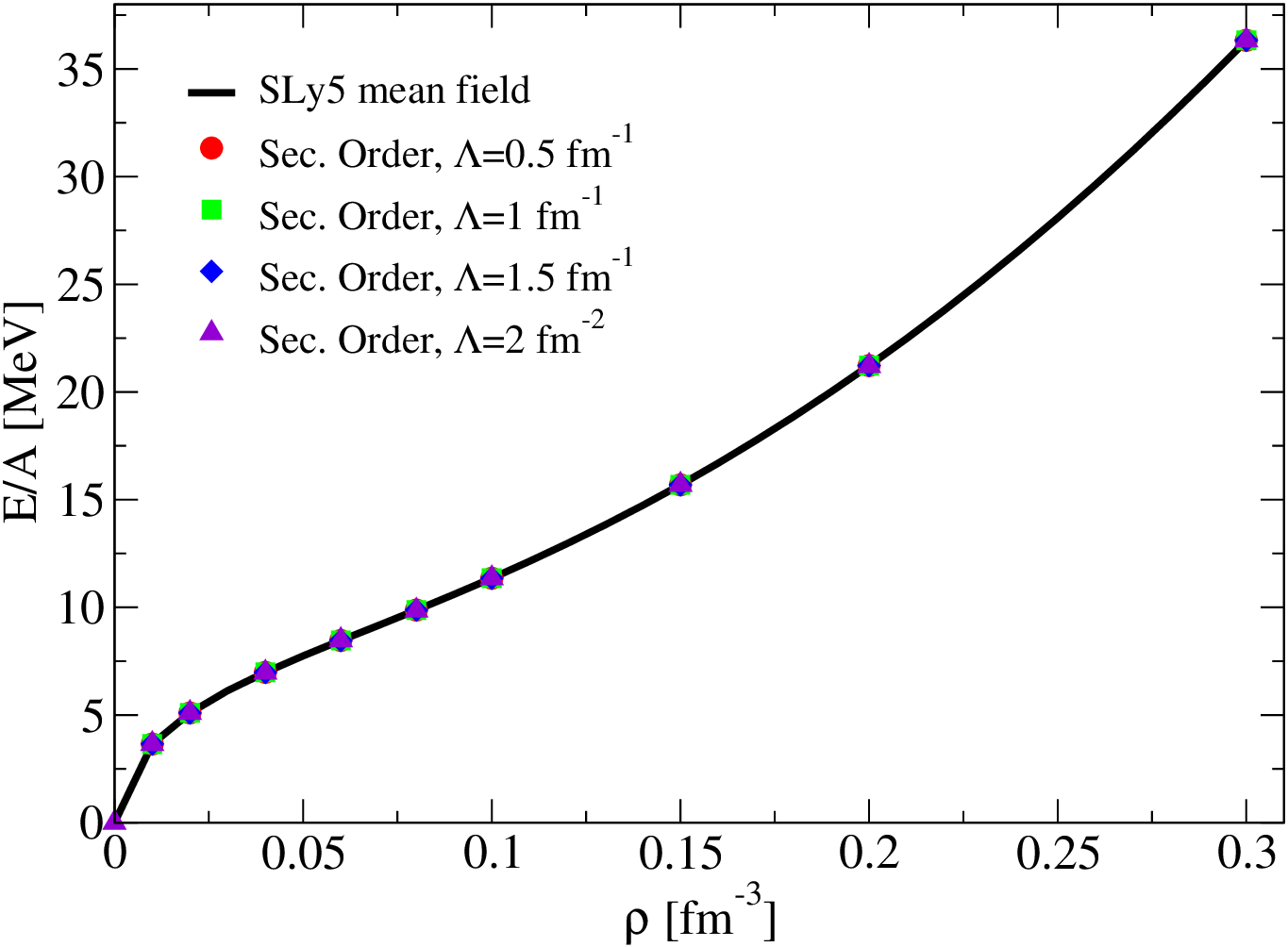}
\caption{(Color online) Refitted second--order EOS for neutron matter
compared with the SLy5--mean--field EOS.}
\label{seven-2}
\end{figure}

We show in Fig. \ref{six} the second--order EOS for pure neutron matter (a)
and the second--order correction (b). In Table II we list the values of the
refitted parameters. Figures \ref{seven-1} and \ref{seven-2} show the
refitted results (differences with respect to the mean--field curve and
absolute values, respectively). The fit is also this time extremely good,
and the $\chi^2$ values are of the order of 10$^{-2}$.

\subsubsection{Asymmetric matter in the case of $\protect\delta=0.5$}

\begin{figure}[tbp]
\includegraphics[scale=0.35]{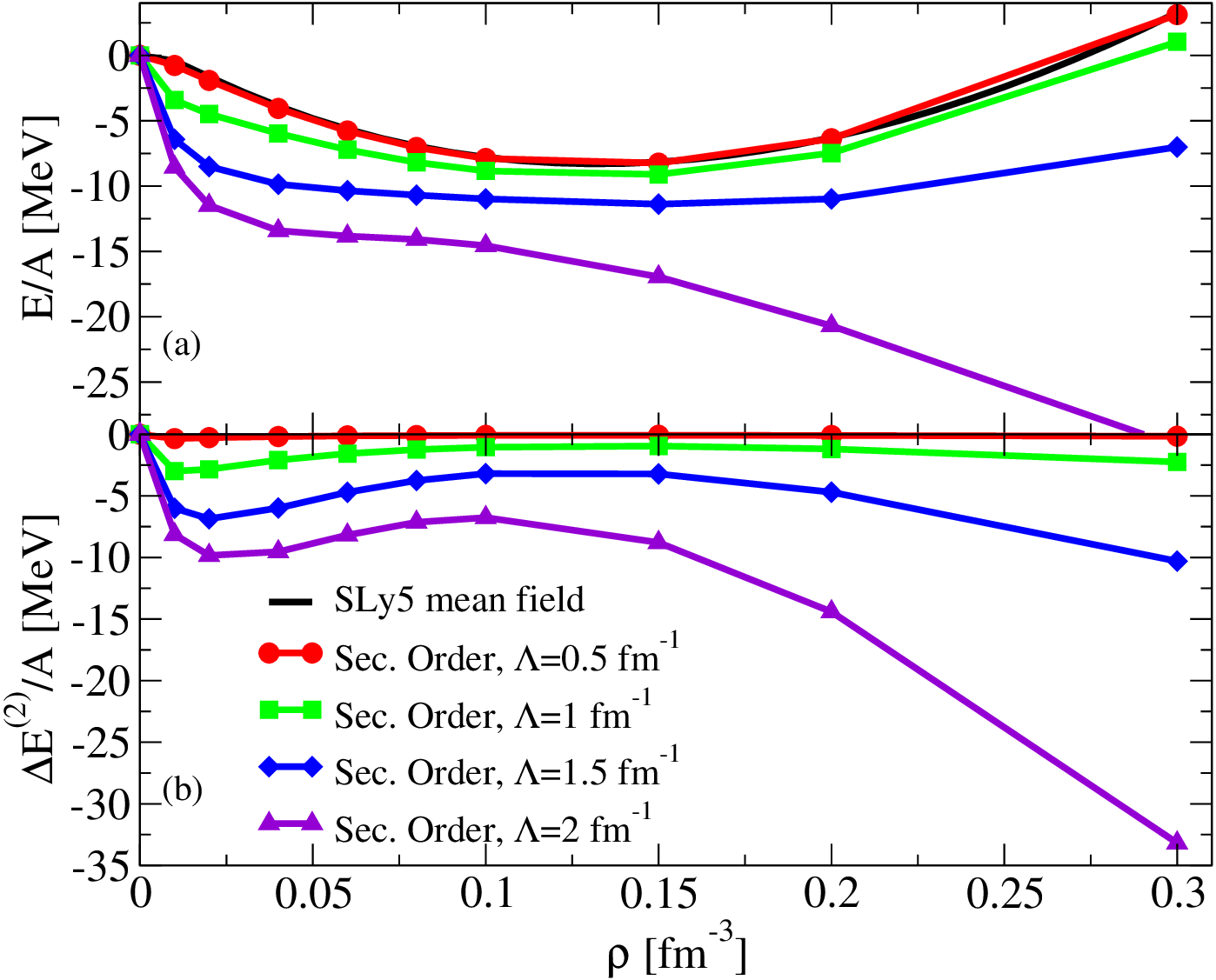}
\caption{(Color online) (a) Second--order EOS of asymmetric matter ($\protect%
\delta=$ 0.5) for several values of the cutoff, calculated with the SLy5
parameters, and SLy5--mean--field EOS; (b) second--order correction.}
\label{eight}
\end{figure}

\begin{figure}[tbp]
\includegraphics[scale=0.35]{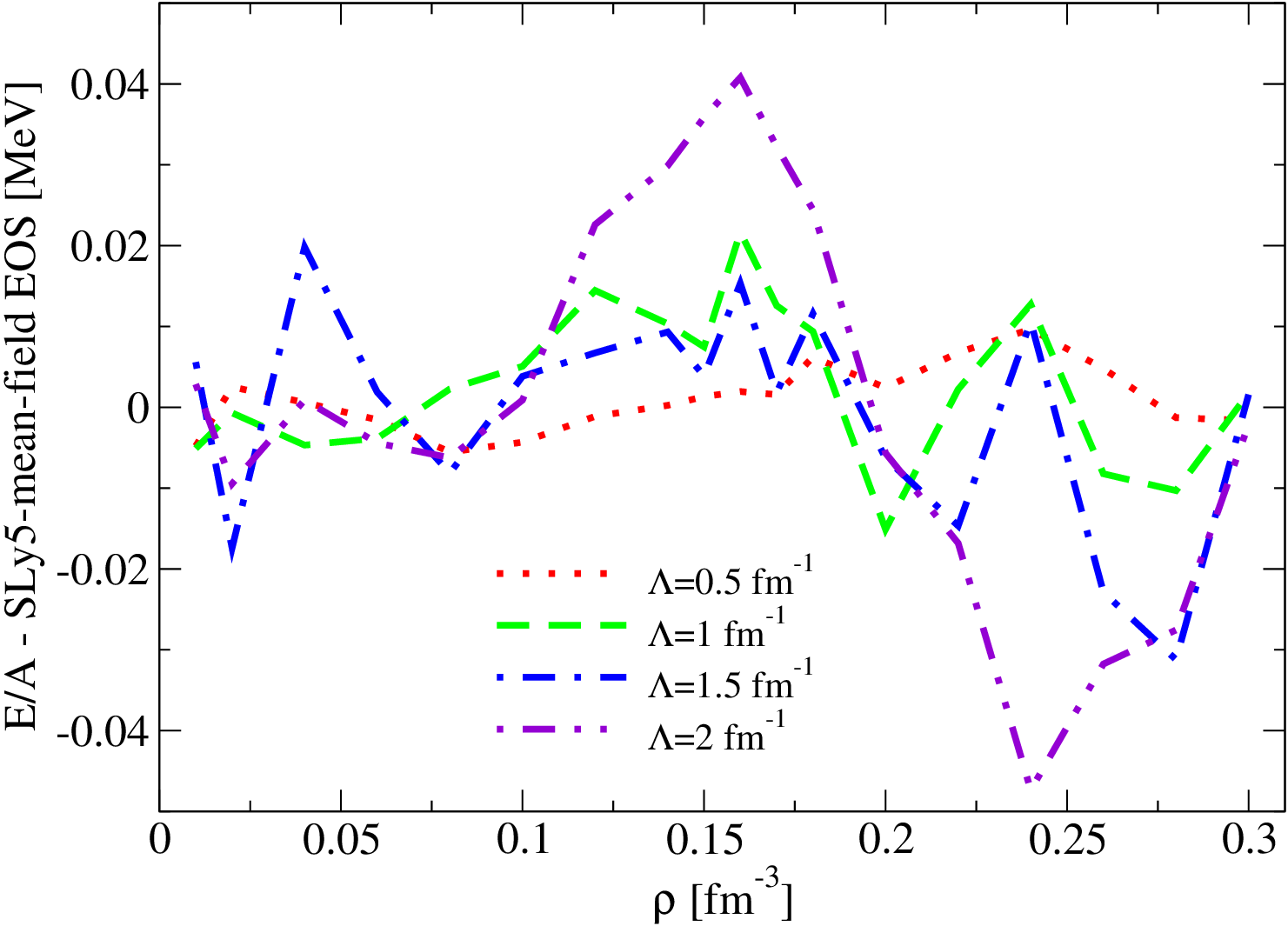}
\caption{(Color online) Difference between the refitted EOS and the
SLy5--mean--field EOS for asymmetric matter ($\protect\delta=$ 0.5).}
\label{nine-1}
\end{figure}

\begin{figure}[tbp]
\includegraphics[scale=0.35]{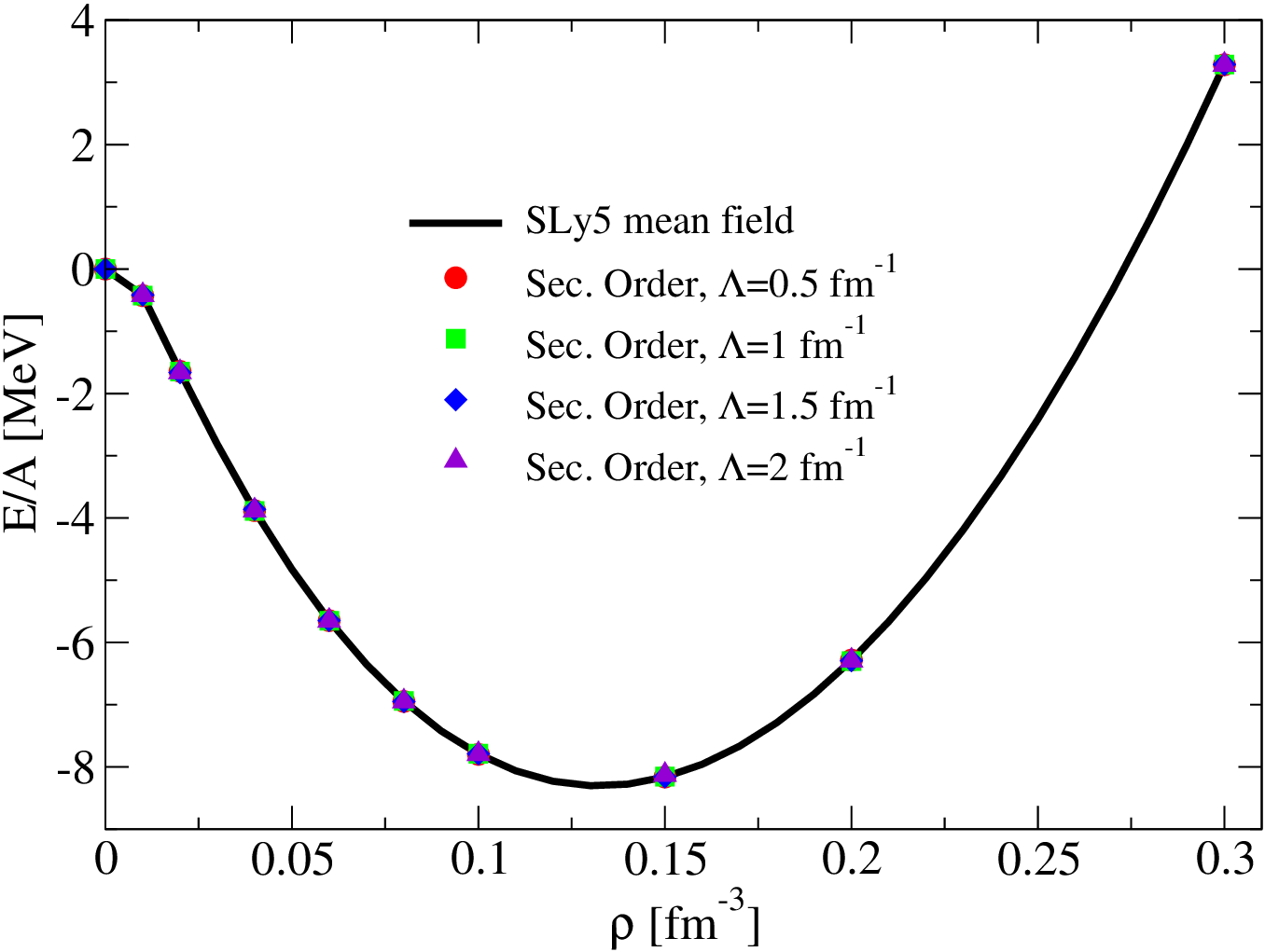}
\caption{(Color online) Refitted second--order EOS for asymmetric matter ($%
\protect\delta=$ 0.5) compared with the SLy5--mean--field EOS.}
\label{nine-2}
\end{figure}

For asymmetric matter we take the illustrative case corresponding to $%
\delta=0.5$. Figure \ref{eight} shows the second--order EOS (a) and the
second--order correction (b). Figures \ref{nine-1} and \ref{nine-2} present
the refitted results shown again as differences with respect to the
benchmark EOS (Fig. \ref{nine-1}) and as absolute values (Fig. \ref{nine-2}%
). Table III contains the refitted parameters and the $\chi^2$ values, which
range from 10$^{-3}$ to 10$^{-1}$, according to the value of the cutoff. The 
$\chi^2$ values are still lower than 1. We can conclude that the quality of
the fit is always extremely good in the three cases of symmetric, neutron,
and asymmetric matter.

\clearpage

\subsubsection{Simultaneous fit of symmetric, neutron, and asymmetric matter
in the case $\protect\delta=0.5$}

\begin{figure}[tbp]
\includegraphics[scale=0.35]{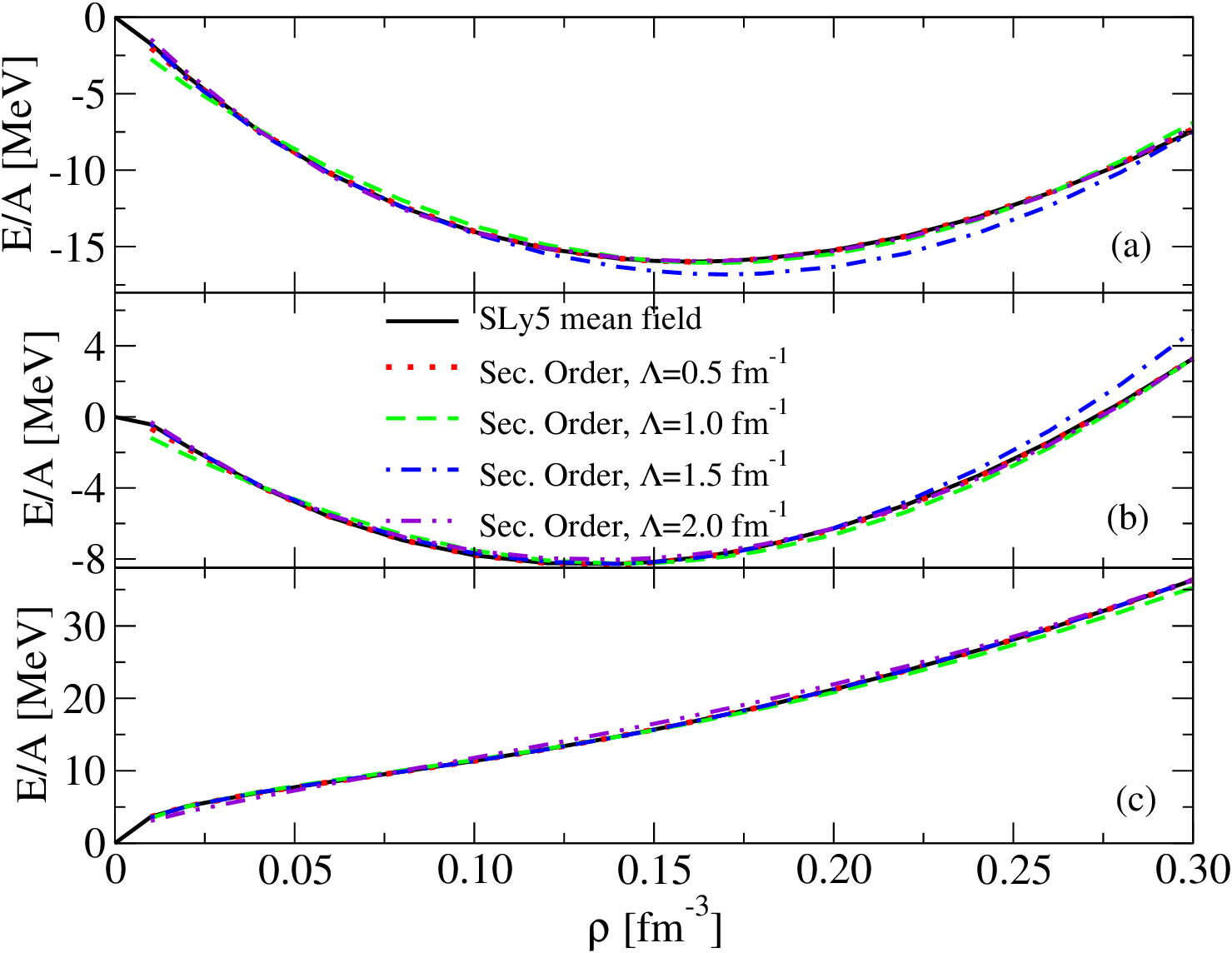}
\caption{(Color online) Second--order EOS's obtained with the simultaneous
fit of symmetric matter (a), asymmetric matter with $\protect\delta=0.5$
(b), and neutron matter (c) compared with the SLy5--mean--field
corresponding curves.}
\label{ten}
\end{figure}

We have then adjusted simultaneously the second--order EOS's of symmetric,
asymmetric ($\delta=0.5$), and neutron matter, as was done in Ref. \cite%
{moghra2012-1}. The obtained curves are shown in Figs. \ref{ten} and \ref%
{eleven} in absolute values and as differences with respect to the benchmark
EOS's, respectively. The values of the adjusted parameters are reported in
Table IV with the associated $\chi^2$ values. We observe that now the
deviations from the benchmark EOS's are larger than in the previous single
fits providing however acceptable EOS's in all cases. The average
discrepancy is less than 2\% for cutoff values of 0.5 and 1 fm$^{-1}$ ($\chi^2$ values 
equal to 0.25 and 3.96, respectively), and
is $\sim$ 3.5\% for cutoff values of 1.5 and 2 fm$^{-1}$ ($\chi^2$ values equal to 
13.9 and 10.7, respectively). The corresponding
pressure and incompressibility modulus are shown in absolute values and as
differences with respect to the mean--field curves in Figs. \ref{inco} and %
\ref{diffeinco}, respectively. The incompressibility modulus at the
saturation point has a maximum deviation from the mean--field value of $\sim$
25 MeV, as indicated by the black arrows in the lower panel of Fig. \ref%
{diffeinco}. Also in this case the pressure is not strictly equal to zero at
saturation. The deviation is now larger than in the previous case (larger
deviations in the saturation density for the different fits).

\begin{figure}[tbp]
\includegraphics[scale=0.35]{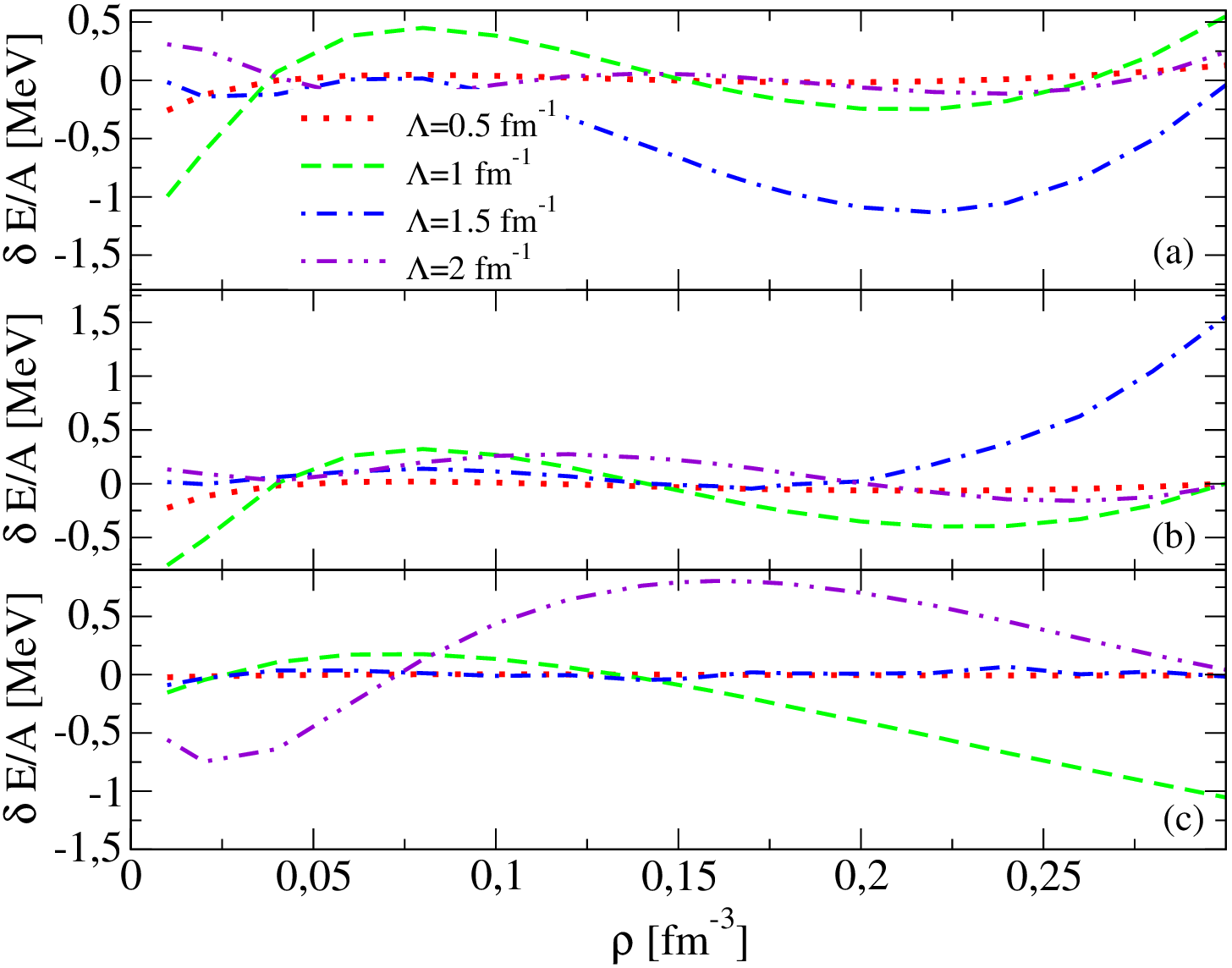}
\caption{(Color online) Difference between the global refitted EOS's and the
SLy5--mean--field EOS's for symmetric (a), $\protect\delta=0.5$ (b) and
neutron (c) matter.}
\label{eleven}
\end{figure}

\begin{figure}[tbp]
\includegraphics[scale=0.35]{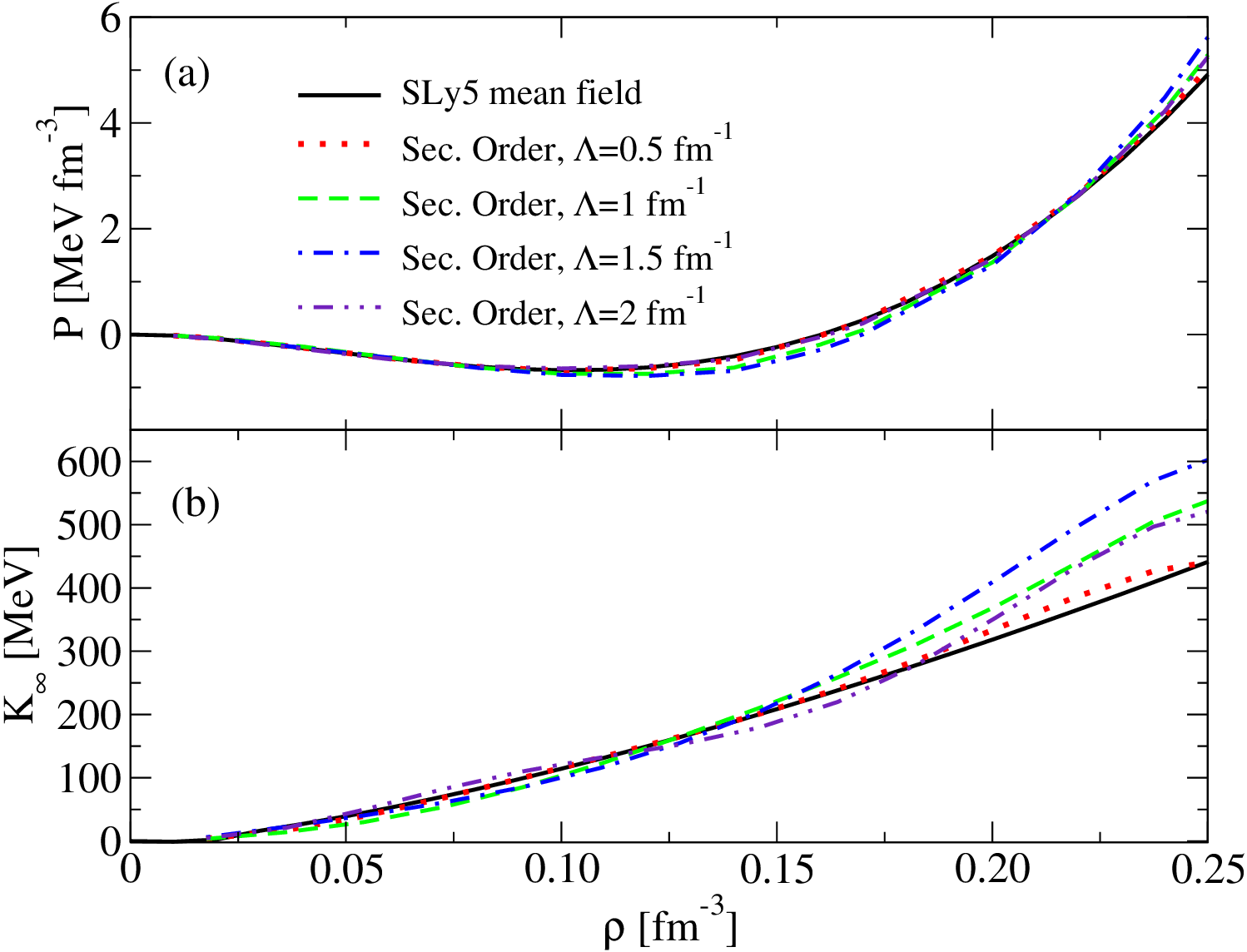}
\caption{(Color online) Pressure (a) and incompressibility modulus (b)
computed with the parameters of the simultaneous fit and compared with the
mean--field curves.}
\label{inco}
\end{figure}

\begin{figure}[tbp]
\includegraphics[scale=0.35]{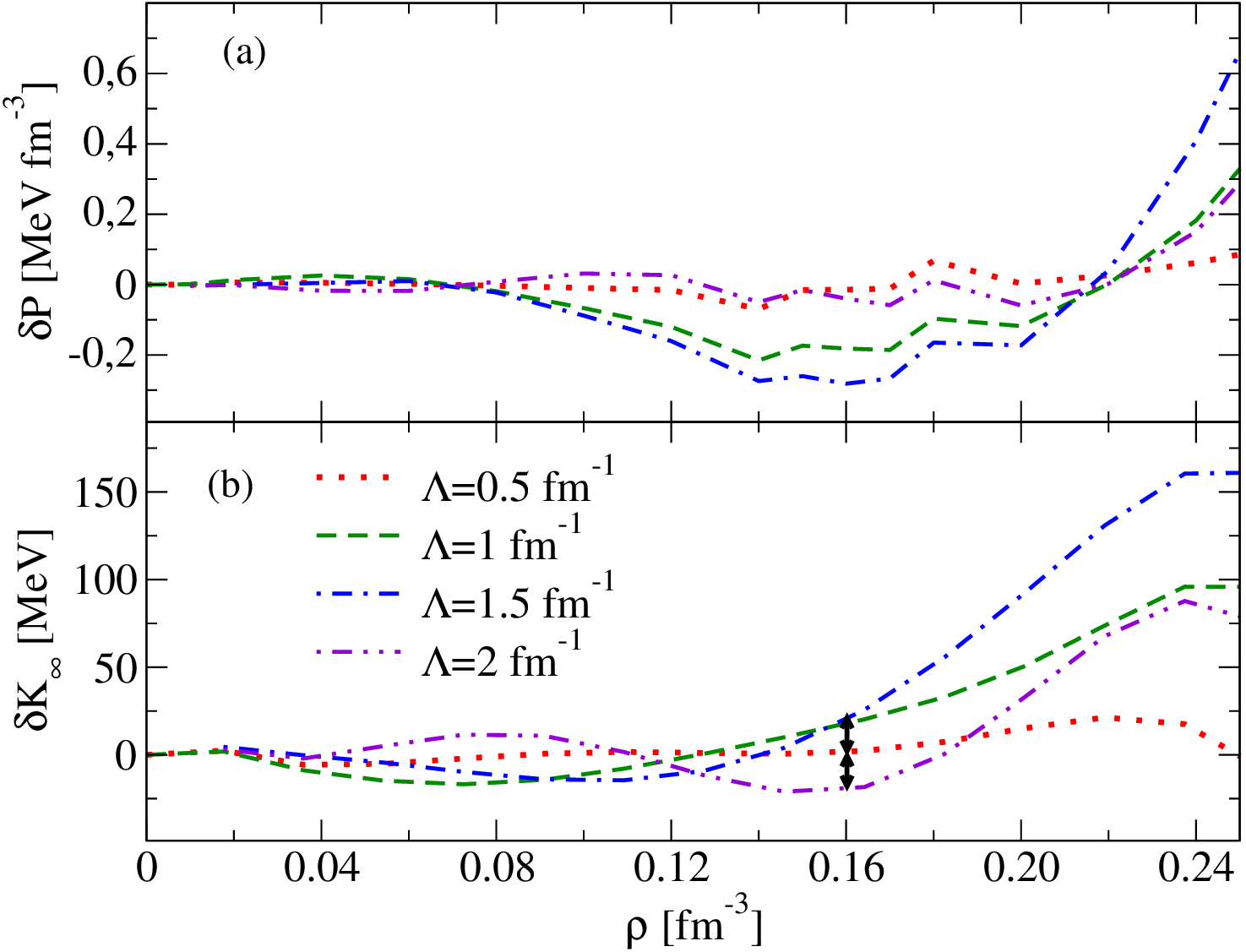}
\caption{(Color online) Differences with respect to the mean--field curves
of the pressure (a) and incompressibility modulus (b), computed with the
parameters of the simultaneous fit.}
\label{diffeinco}
\end{figure}
\clearpage

\section{Dimensional regularization}

In the present Section, we report the revised results, with respect to Ref. 
\cite{moghra2012-2}, concerning the application of the dimensional
regularization to the second--order integrals and the extraction of the
corresponding finite contributions in the EOS's of symmetric, neutron, and
asymmetric matter, for the case $\delta=0.5$. This regularization technique,
that was originally introduced for the electroweak theory \cite%
{hooft1,hooft2,bollini}, is based on the solution of the divergent integrals
with the use a continuous parameter $d$ which replaces their integer
dimension. After the evaluation of the integral, the parameter $d$ returns
to the integer value.

In Ref. \cite{moghra2012-2}, the first analyzed case was the simple
model $t_0-t_3$ applied to symmetric matter (the case analogous to that of
Ref. \cite{moghraprl}). This was correctly treated and the corresponding
results are reported in Sects. II.A and III.A and up to Fig. 2 of Ref. \cite%
{moghra2012-2}.

For the other cases (Skyrme interaction containing all terms except the
tensor and the spin--orbit parts, and the treatment of also asymmetric and
neutron matter) the same corrections done in the previous section have to be
performed in the evaluation of the second--order contribution. We provide in
what follows the analytical expressions of the second--order contributions
for symmetric and neutron matter. We analyze the results for symmetric,
neutron, and asymmetric matter in the case $\delta=0.5$. The case of
asymmetric matter is not derived analytically, but the finite part of this
EOS is extracted from the numerical Monte Carlo calculation.

Also in the present section we employ the approximation $m^*=m$ and we neglect 
the rearrangement terms at second order. These approximations will be abandoned in Sect. IV.

\subsection{Symmetric matter}

\begin{figure}[tbp]
\includegraphics[scale=0.35]{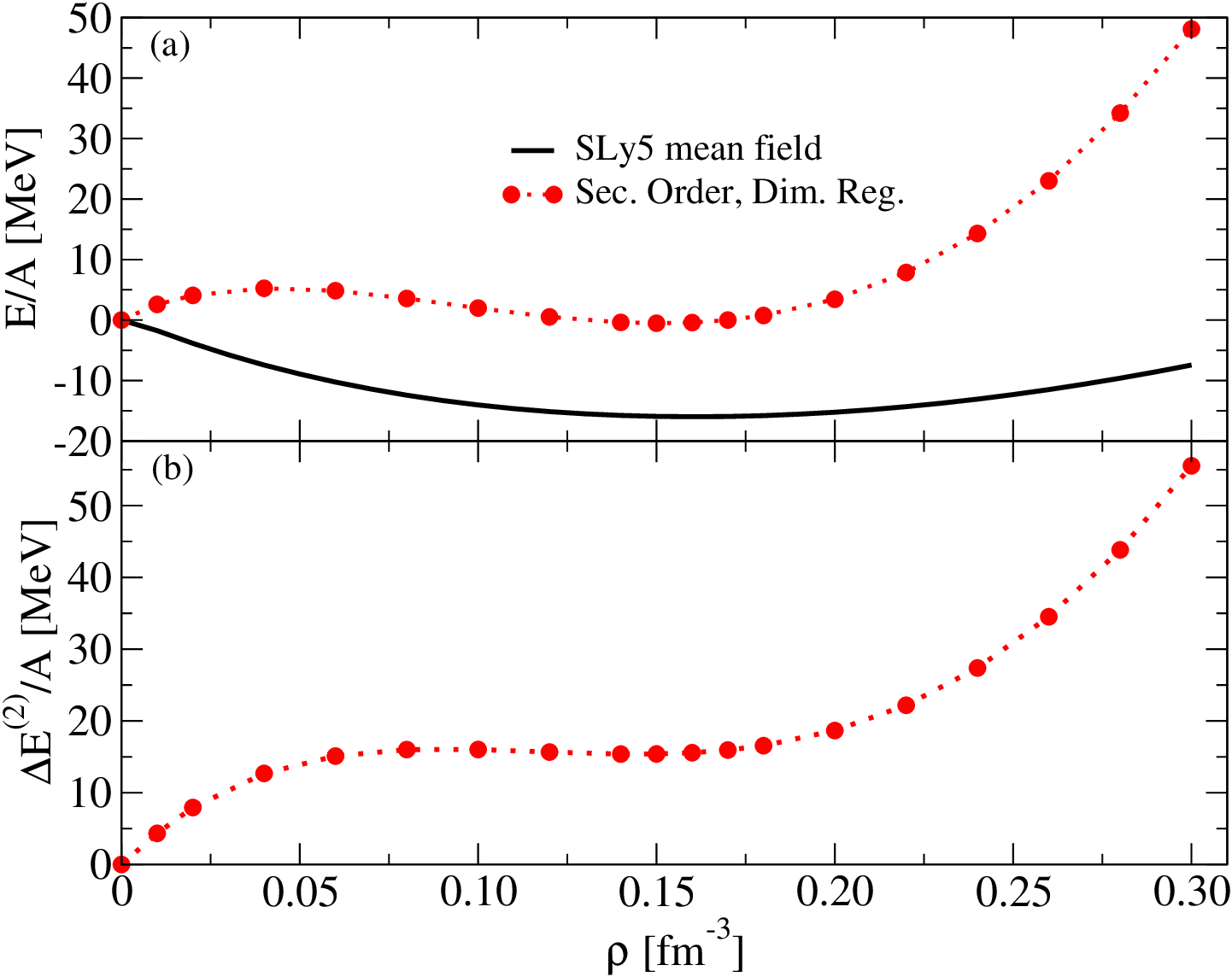}
\caption{(Color online) (a) Dimensional--regularized second--order EOS for
symmetric matter compared with the corresponding SLy5--mean--field EOS. (b) Second--order correction.}
\label{symmdr}
\end{figure}

\begin{figure}[tbp]
\includegraphics[scale=0.35]{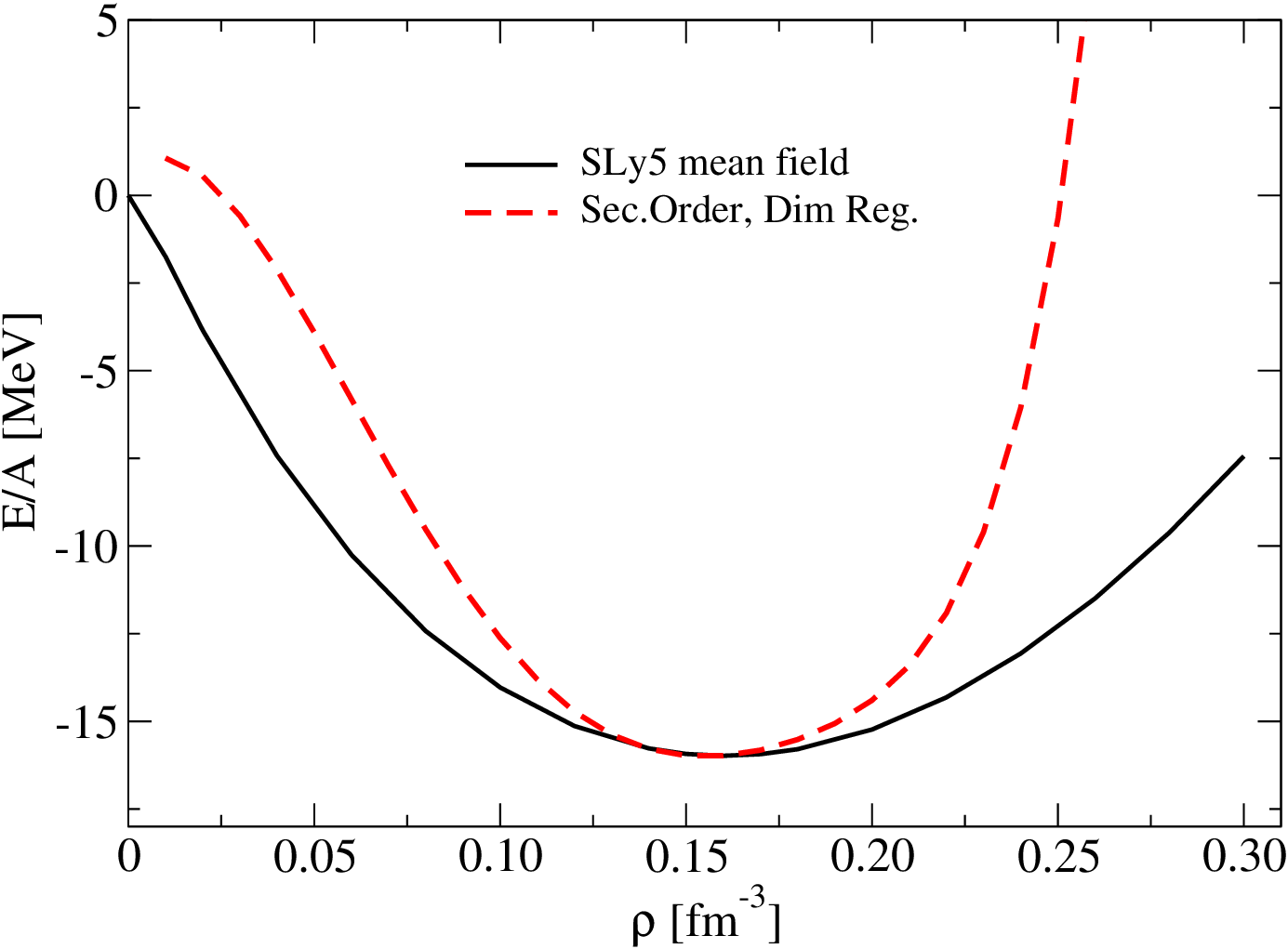}
\caption{(Color online) Refitted dimensional--regularized second--order EOS
for symmetric matter compared with the corresponding SLy5--mean--field EOS.}
\label{symmdrfit}
\end{figure}

Starting from the Skyrme interaction of Eq. (\ref{sk}), the
dimensional--regularized second--order result for symmetric matter is
written as 
\begin{eqnarray}
\frac{E_{sym}^{(2)F}}{A}&=&\frac{mk_{F}^{4}}{\hbar^2\pi ^{4}} ( \frac{3}{560}
(11-2 ln[2]) \widetilde{T}_{03}^{2} + \frac{1}{1260} (167 -24 ln[2]) k_F^2 
\widetilde{T}_{03} \widetilde{T}_{1} +\frac{1}{55440} (4943-564 ln[2]) k_F^4 
\widetilde{T}_{1}^2  \notag \\
&+& \frac{1}{73920}(1033 - 156 ln[2]) k_F^4 \widetilde{T}_{2}^{2} ),
\label{finitesy}
\end{eqnarray}
where ``$F$'' stands for finite part. 
The second--order EOS compared with the SLy5--mean--field EOS is displayed
in Fig. \ref{symmdr}. The refitted curve is plotted in Fig. \ref{symmdrfit}
and the corresponding parameters are written in Table V ($\delta=0$). 

We could correctly reproduce the saturation region only by  
restricting the fit of the parameters in a narrow region of densities close to the saturation point. 
The EOS is however poorly described in the other density regions, as can be seen in Fig. 
\ref{symmdrfit}; this implies that the incompressibility modulus is not well reproduced (368 MeV). 
An improvement is however seen with respect to the adjustment presented  
in Ref. \cite{kaiser}, where the equilibrium point was completely missed. This improvement is due to the density--dependent term which allows us at least to 
shift the equilibrium point to its correct value. We do not report for this case the $\chi^2$ value because the reference points for the fit are 
taken only close to the saturation point and such $\chi^2$ value would thus not be comparable with those obtained from the previous fits where the reference points were distributed in the whole region of densities.


\subsection{Neutron matter}

\begin{figure}[tbp]
\includegraphics[scale=0.35]{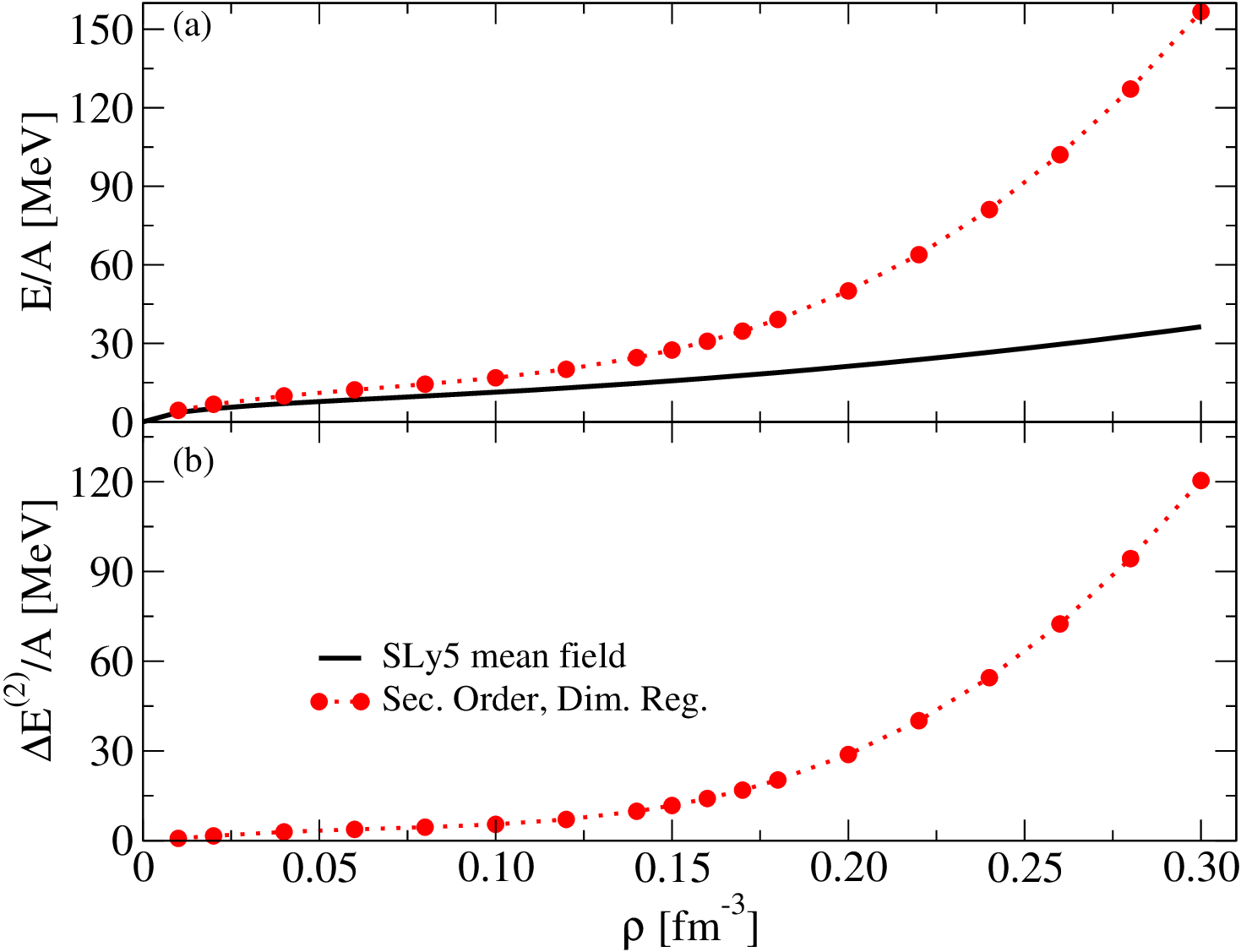}
\caption{(Color online) (a) Dimensional regularized second--order EOS for
neutron matter compared with the corresponding SLy5--mean--field EOS. (b) Second--order 
correction.}
\label{neutrdr}
\end{figure}

\begin{figure}[tbp]
\includegraphics[scale=0.35]{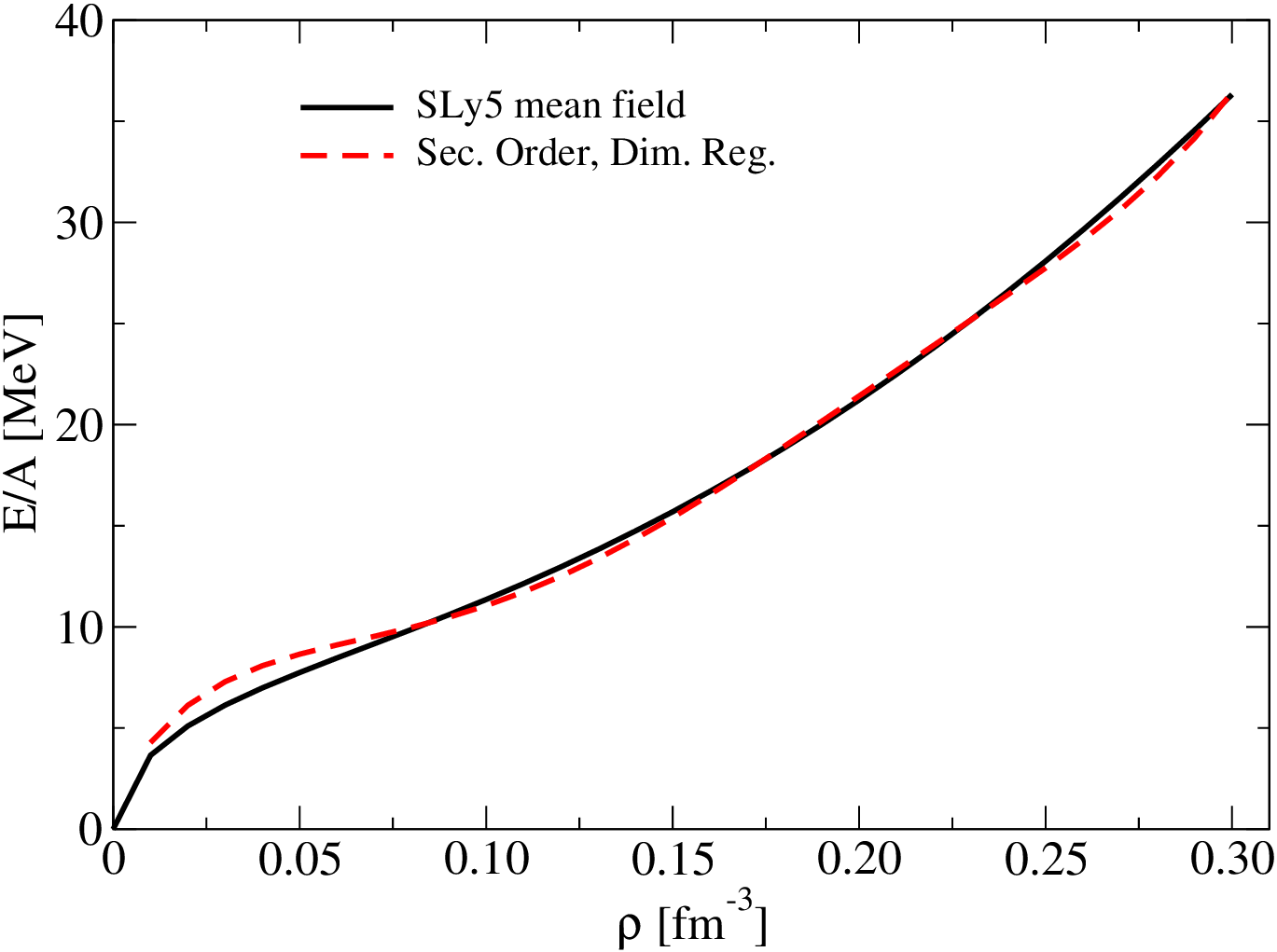}
\caption{(Color online) Best fit for dimensional regularized second--order
EOS of neutron matter compared with the SLy5--mean--field EOS.}
\label{twelve}
\end{figure}

The analytical expression for the second--order contribution is 
\begin{eqnarray}
\frac{E_{neutr}^{(2)F}}{A}&=&\frac{mk_{N}^{4}}{\hbar^2\pi ^{4}} ( \frac{1}{280}
(11-2 ln[2]) T_{03}^{2} + \frac{1}{1890} (167 -24 ln[2]) k_N^2 T_{03} T_{1} +%
\frac{1}{83160} (4943-564 ln[2]) k_N^4 T_{1}^2  \notag \\
&+& \frac{1}{110880}(1033 - 156 ln[2]) k_N^4 T_{2}^{2} ).  \label{finitene}
\end{eqnarray}
The corresponding EOS is plotted in Fig. \ref{neutrdr} and compared with the
corresponding SLy5--mean--field EOS. Figure \ref{twelve} shows the curve
obtained with the adjusted parameters and Table V presents the associated
parameters ($\delta=1$).In this case, the 9 reference points chosen 
for the fit are distributed in the whole region of densities. The $\chi^2$ value is $\sim$ 91, 
indicating an average deviation from the reference points of about 9.5 \%. The quality of the fit 
is lower than that obtained for neutron matter in the case of the cutoff regularization but is 
still reasonably good.

\subsection{Asymmetric matter}

\begin{figure}[tbp]
\includegraphics[scale=0.35]{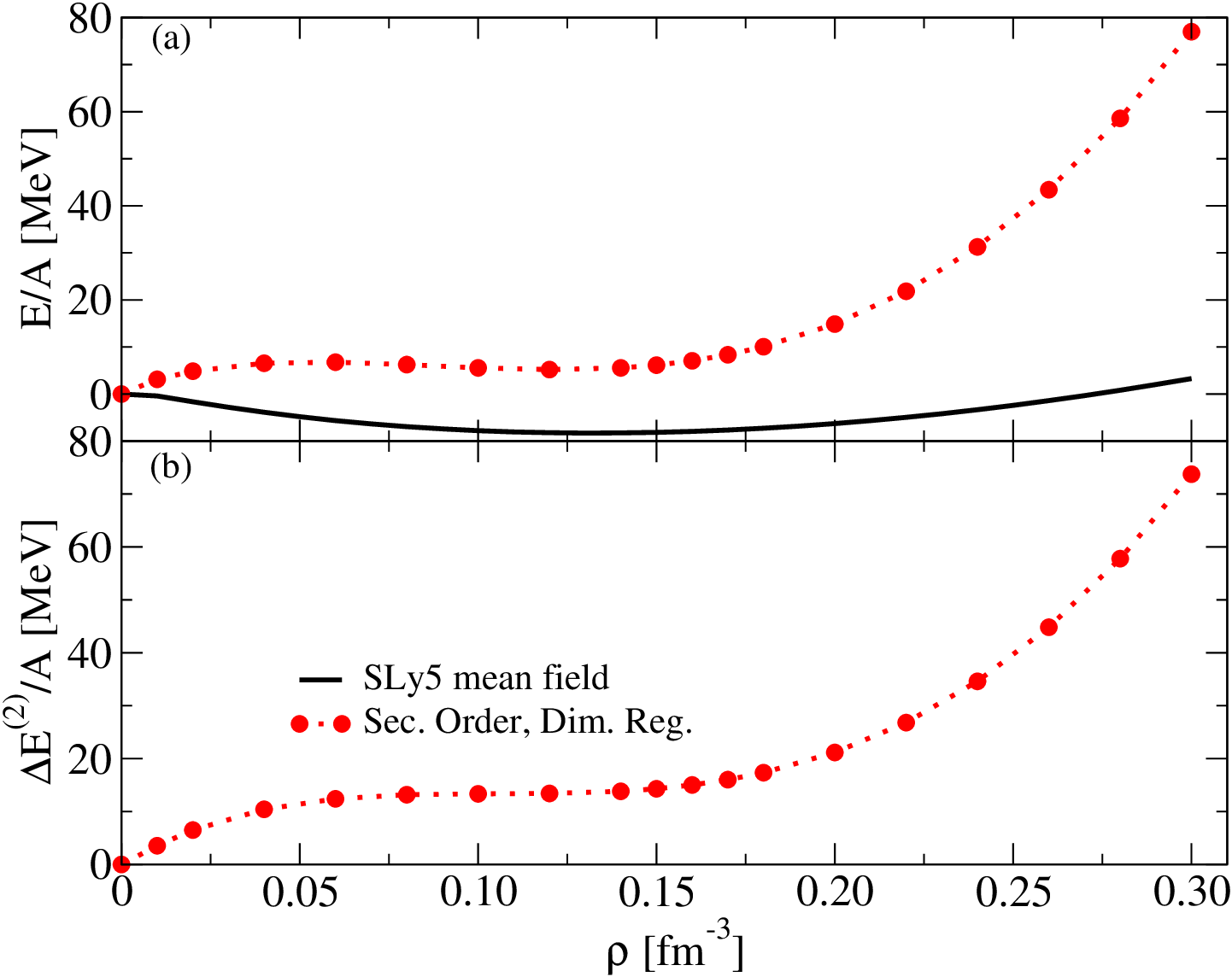}
\caption{(Color online) (a) Dimensional--regularized second--order EOS for
asymmetric matter in the case $\protect\delta=0.5$ compared with the
corresponding SLy5--mean--field EOS. (b) Second--order correction.}
\label{asymmdr}
\end{figure}

The second--order EOS of asymmetric matter (for the case $\delta=0.5$) and
the refitted curve are plotted respectively in Figs. \ref{asymmdr} and \ref%
{thirteen}. The parameters are listed in Table V. 
Also in this case, as for symmetric matter, the equilibrium region can be described only by taking a narrow region 
of densities around the minimum to perform the adjustment of the parameters. The EOS is clearly very poorly
 described in the other density regions. 

We may conclude that, when  
dimensional regularization is used, the fit of the parameters has a global 
good quality 
only for the case of pure neutron matter. We will see however in next section that 
results are considerably improved for symmetric matter when the rearrangement terms are taken into account.

\begin{figure}[tbp]
\includegraphics[scale=0.35]{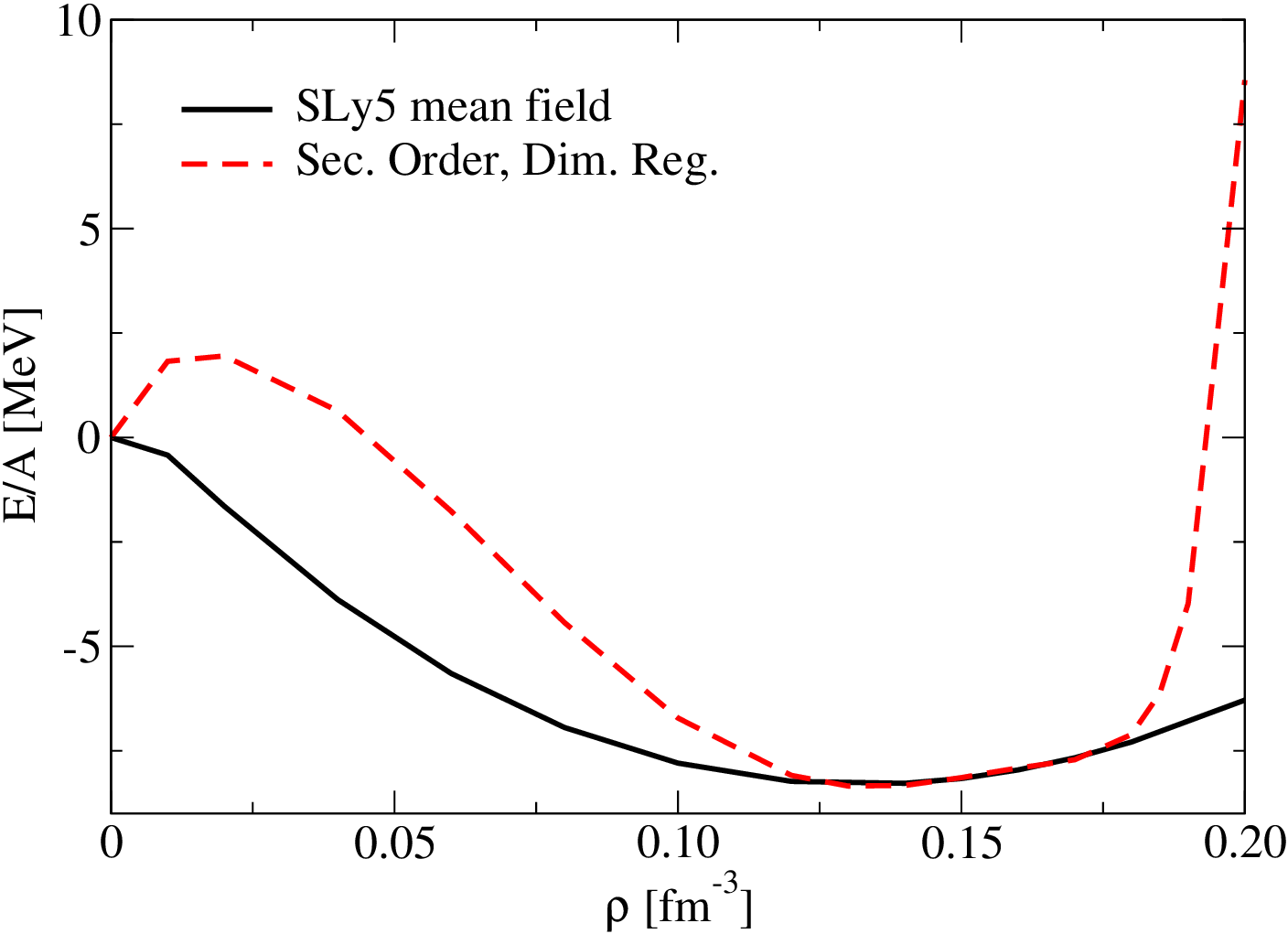}
\caption{(Color online) Best fit for dimensional--regularized second--order
EOS for asymmetric matter ($\protect\delta=$ 0.5) matter compared with the
SLy5--mean--field EOS. }
\label{thirteen}
\end{figure}

\section{Effective mass and rearrangement terms}
We worked so far by using two approximations: We have approximated the effective mass with the bare mass and we have neglected the second--order rearrangement terms generated by the density--dependent part of the 
interaction. In the present section, we include first an effective mass in the computation of the second--order EOS for the cases of 
symmetric and pure neutron matter. Then, we also include in these cases the corresponding rearrangement terms.

\begin{figure}[tbp]
\includegraphics[scale=0.35]{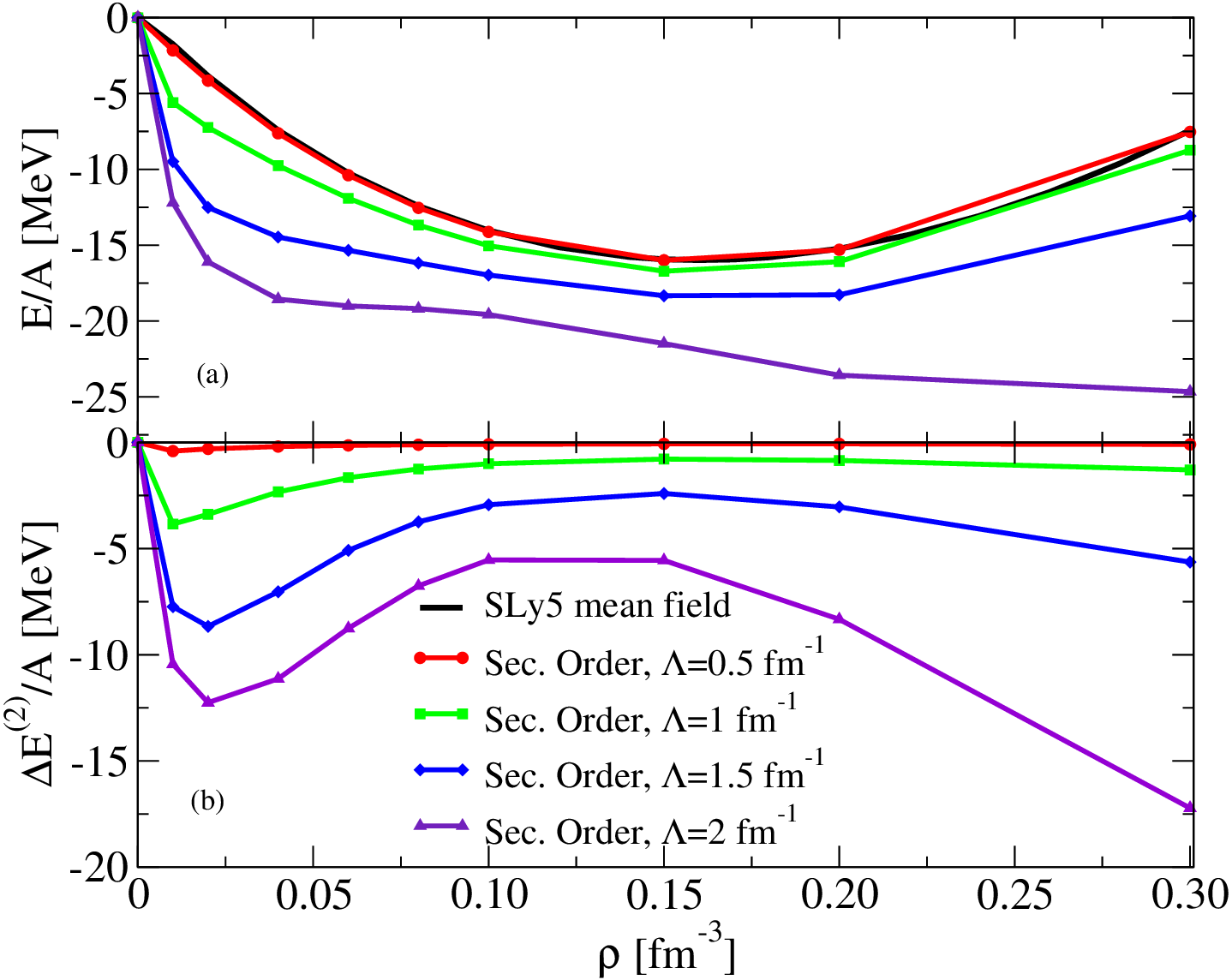}
\caption{(Color online) Same as Fig. \ref{two} but with an effective mass computed within the mean--field approximation.}
\label{mstarsym}
\end{figure}

\begin{figure}[tbp]
\includegraphics[scale=0.35]{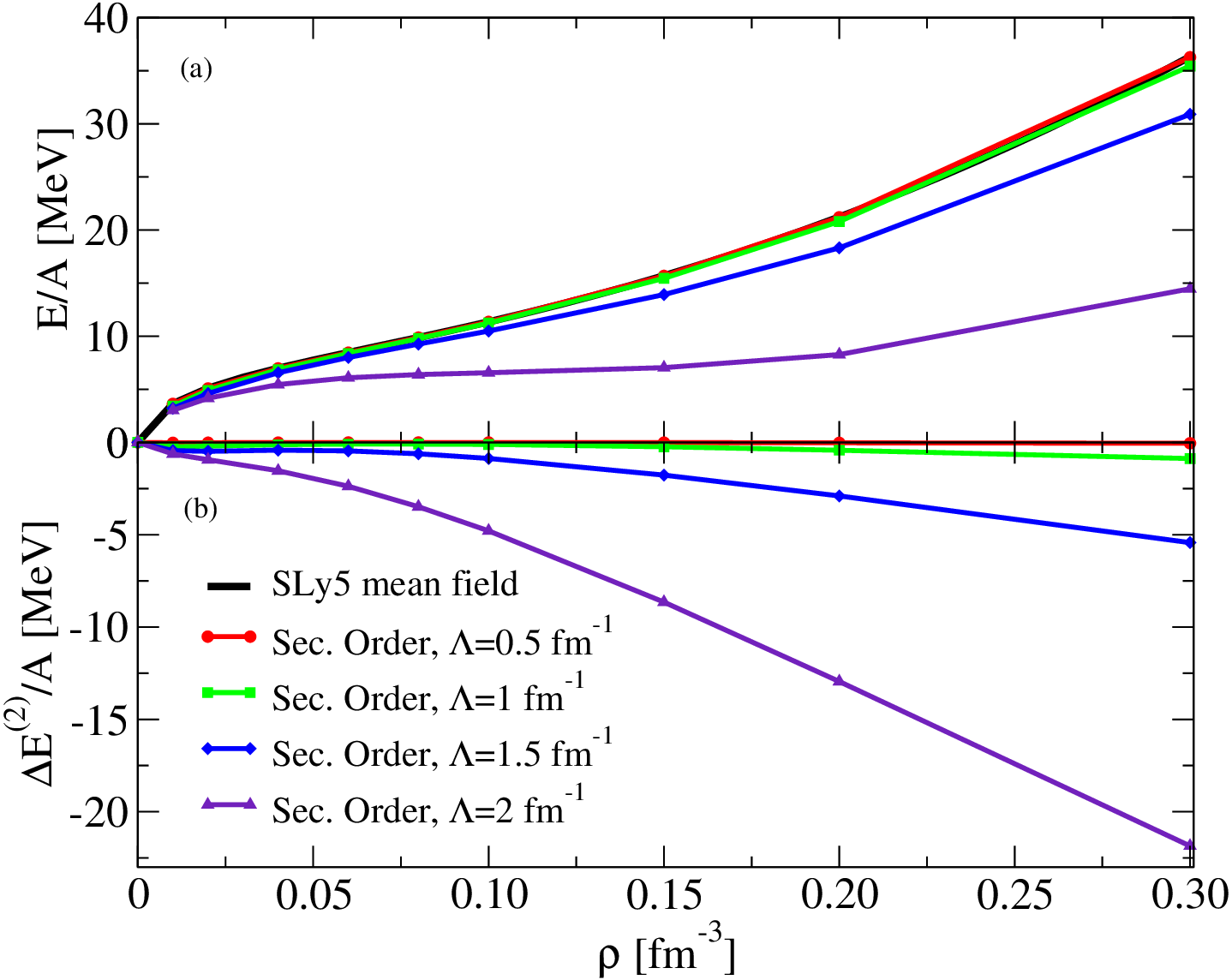}
\caption{(Color online) Same as Fig. \ref{six} but with an effective mass computed within the mean--field approximation.}
\label{mstarneut}
\end{figure}

We use the mean--field approximation for the effective mass, where only the velocity--dependent terms contribute. Fos symmetric matter, one has 
\begin{equation}
\frac{m^*_{S}}{m}=\left(1+\frac{m}{8\hbar^2}\rho \Theta_S \right)^{-1},
\label{msym}
\end{equation}
whereas for neutron matter one has 
\begin{equation}
\frac{m^*_{N}}{m}=\left(1+\frac{m}{4\hbar^2}\rho \Theta_S - \frac{m}{4\hbar^2}\rho \Theta_V \right)^{-1}.
\label{mneu}
\end{equation}
By using $m^*$ instead of $m$ in Eq. (\ref{prop}), new EOS's for symmetric and pure neutron matter are obtained. 
In the case of cutoff regularization, they are evaluated  by 
multiplying Eqs. (\ref{symml0}), (\ref{symml1}), and (\ref{neutl0}), (\ref{neutl1}) by Eqs. (\ref{msym}) and (\ref{mneu}), respectively. 
In the case of dimensional regularization, they are obtained by 
multiplying Eqs. (\ref{finitesy}) and (\ref{finitene}) by Eqs. (\ref{msym}) and (\ref{mneu}), respectively.
The corresponding 
curves evaluated with the SLy5 parameters show a density--dependent rescaling effect, as obviously expected. 
This may be observed in the illustrative case of cutoff regularization: 
Figures \ref{mstarsym} and \ref{mstarneut} describe the same quantities as Figs. 
 \ref{two} and \ref{six}, but with $m^*\ne m$. 

\begin{figure}[tbp]
\includegraphics[scale=0.35]{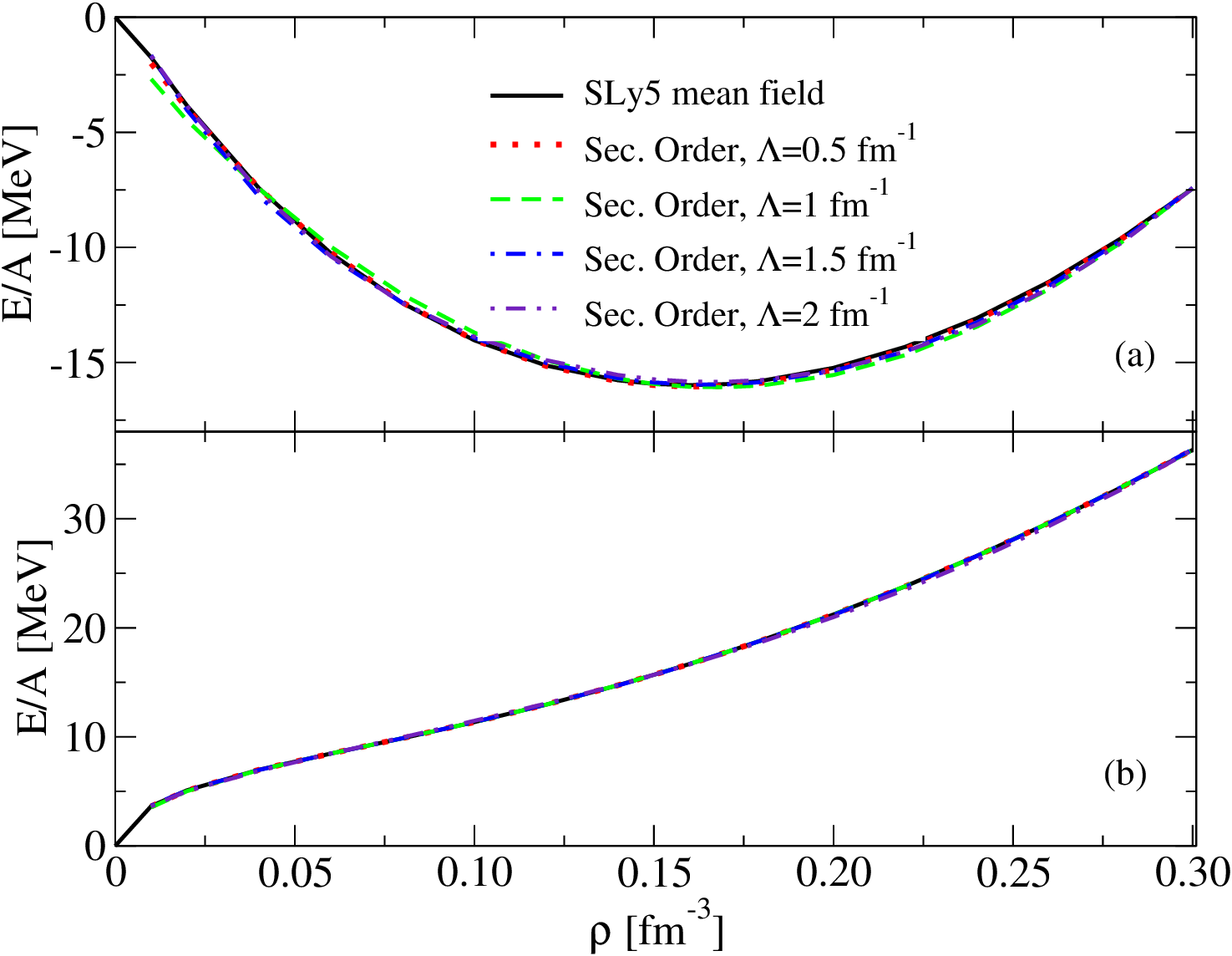}
\caption{(Color online) Second--order EOS's for symmetric (a)  and pure neutron matter (b) adjusted on the SLy5--mean--field EOS's, with an effective mass equal to the 
mean--field effective mass, for different values of the cutoff.}
\label{mstarglobal}
\end{figure}

Fits of parameters may be done for symmetric and pure neutron matter and we provide here the illustrative results obtained in the 
case of the cutoff regularization. We show in Fig. \ref{mstarglobal} the curves corresponding to  
a global fit done on the benchmark SLy5 EOS's by including simultaneously symmetric and pure neutron matter. The obtained parameters are listed in Table VI. 
The incompressibility modulus ranges from 202 to 238 MeV, according to the different cutoff values. The quality of the fit is globally very good with a maximum average deviation from the reference curve of $\sim$ 3\%. 
We mention that, in the case of dimensional regularization, we found that 
the inclusion of an effective mass is not sufficient to improve the quality of the fit 
of the second--order EOS for symmetric matter, which remains similar to that shown in Fig. \ref{symmdrfit}.

We discuss now the  rearrangement terms. 
As already anticipated in Sect. I, Ref. \cite{car} pointed out that the square of the interaction entering in the computation of the second--order energy correction coincides with the square of the RPA $B$ matrix, which means that the rearrangement terms have to be computed by using the second derivative of the Hartree-Fock energy functional, as done in RPA.
By following such procedure and using the Landau parameters computed with the Skyrme force for symmetric and 
pure neutron matter \cite{pastore}, 
 the combinations of parameters containing the $t_3$ part in Eqs. (\ref{tnomsymm}) and (\ref{tnomneutr}) may be replaced by

\begin{eqnarray}
\nonumber
\widetilde{T^{R}}_{03}^{2} &=&\left[ t_{0}(1-x_{0})+\frac{1}{6}%
t_{3}(1-x_{3})\rho ^{\alpha}+\frac{1}{32}t_3 \rho^{\alpha}\alpha(3+\alpha)\right] ^{2} \\
\nonumber 
&+&\left[ t_{0}(1+x_{0})+\frac{1}{6}%
t_{3}(1+x_{3})\rho ^{\alpha }+\frac{1}{32}t_3 \rho^{\alpha}\alpha(3+\alpha)\right] ^{2},    \\
\nonumber
\widetilde{T^{R}}_{03}\widetilde{T^R}_{1} &=&\frac{t_{1}}{2}\left[ [t_{0}(1-x_{0})+%
\frac{1}{6}t_{3}(1-x_{3})\rho ^{\alpha }+\frac{1}{32}t_3 \rho^{\alpha}\alpha(3+\alpha)](1-x_{1})\right] \\ 
\nonumber 
&+&\frac{t_{1}}{2}\left[[t_{0}(1+x_{0})+\frac{1}{6%
}t_{3}(1+x_{3})\rho ^{\alpha } +\frac{1}{32}t_3 \rho^{\alpha}\alpha(3+\alpha) ](1+x_{1})\right],   \\ 
T^R_{03} &=&t_{0}(1-x_{0})+\frac{1}{6}t_{3}(1-x_{3})\rho ^{\alpha }  
+ \frac{1}{48}t_3 \rho^{\alpha}\alpha(\alpha+3)(1-x_3),
\end{eqnarray}%
where ``R'' indicates the inclusion of the rearrangement terms.
 
\begin{figure}[tbp]
\includegraphics[scale=0.35]{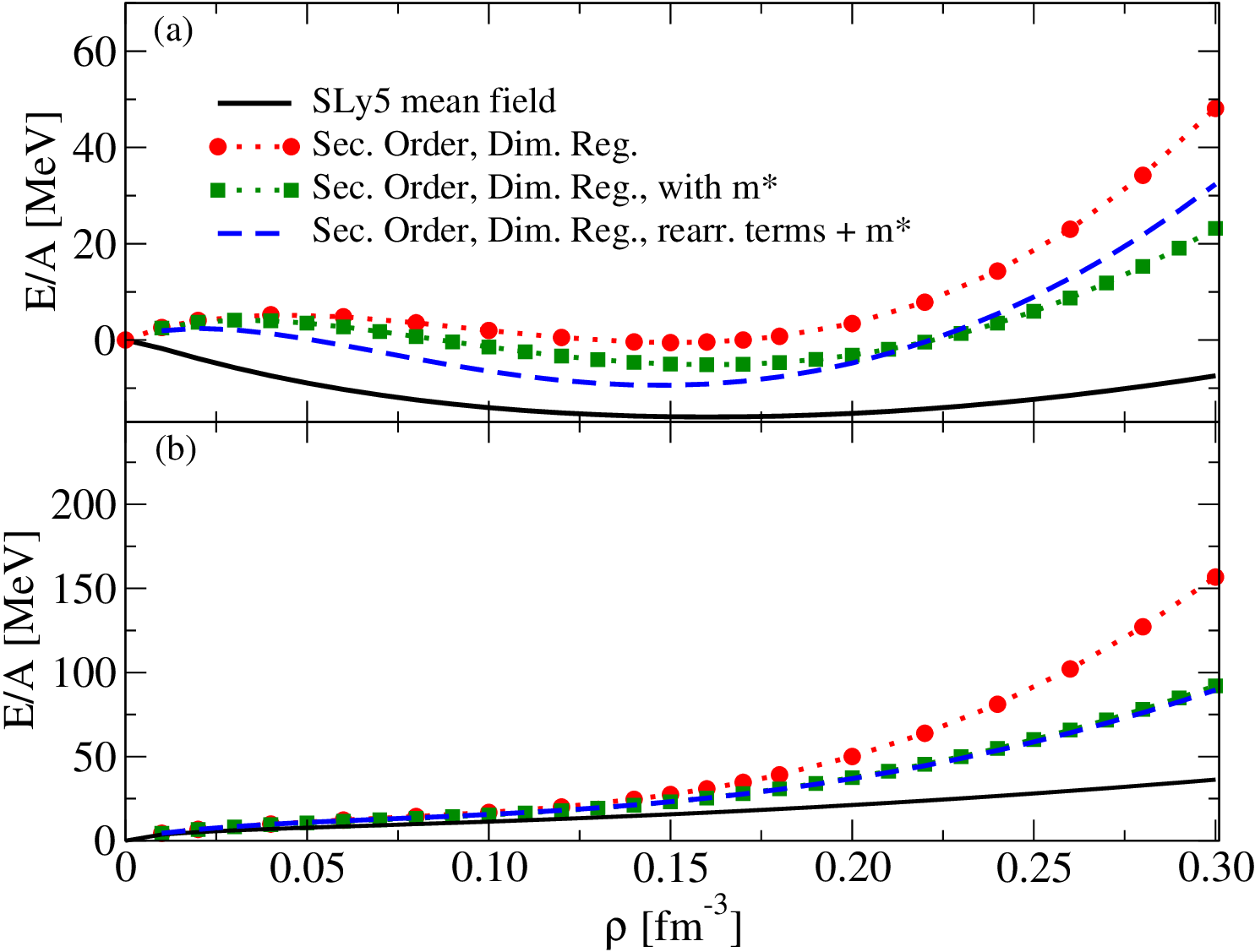}
\caption{(Color online) Second--order EOS's for symmetric (a)  and pure neutron matter (b) compared with the SLy5--mean--field EOS's, calculated with the 
SLy5 parameters for the case where $m^*= m$ and rearrangement terms are neglected (red circles and dotted line), for the case where a mean--field effective mass is used and rearrangement terms are neglected (gree squares and dotted line), and for the case where a mean--field effective mass is used and rearrangement terms are included (blue dashed line).}
\label{rearreffect}
\end{figure}

Figure \ref{rearreffect} shows, as an illustration, how the dimensional--regularized second--order 
EOS's (computed with the SLy5 parameters) 
are modified by the inclusion of the rearrangement terms. 
One observes that, for the case of neutron matter, the inclusion of rearrangement terms in the EOS 
has a very weak effect compared to the much more important effect coming from 
the inclusion of the effective mass. On the other side, for the case of 
symmetric matter, rearrangement terms modify the curve. 
Such modification, with respect to the case where rearrangement terms were omitted,
allows us to obtain a much better refitted dimensional--regularized EOS for symmetric matter, as will be shown below.

\begin{figure}[tbp]
\includegraphics[scale=0.35]{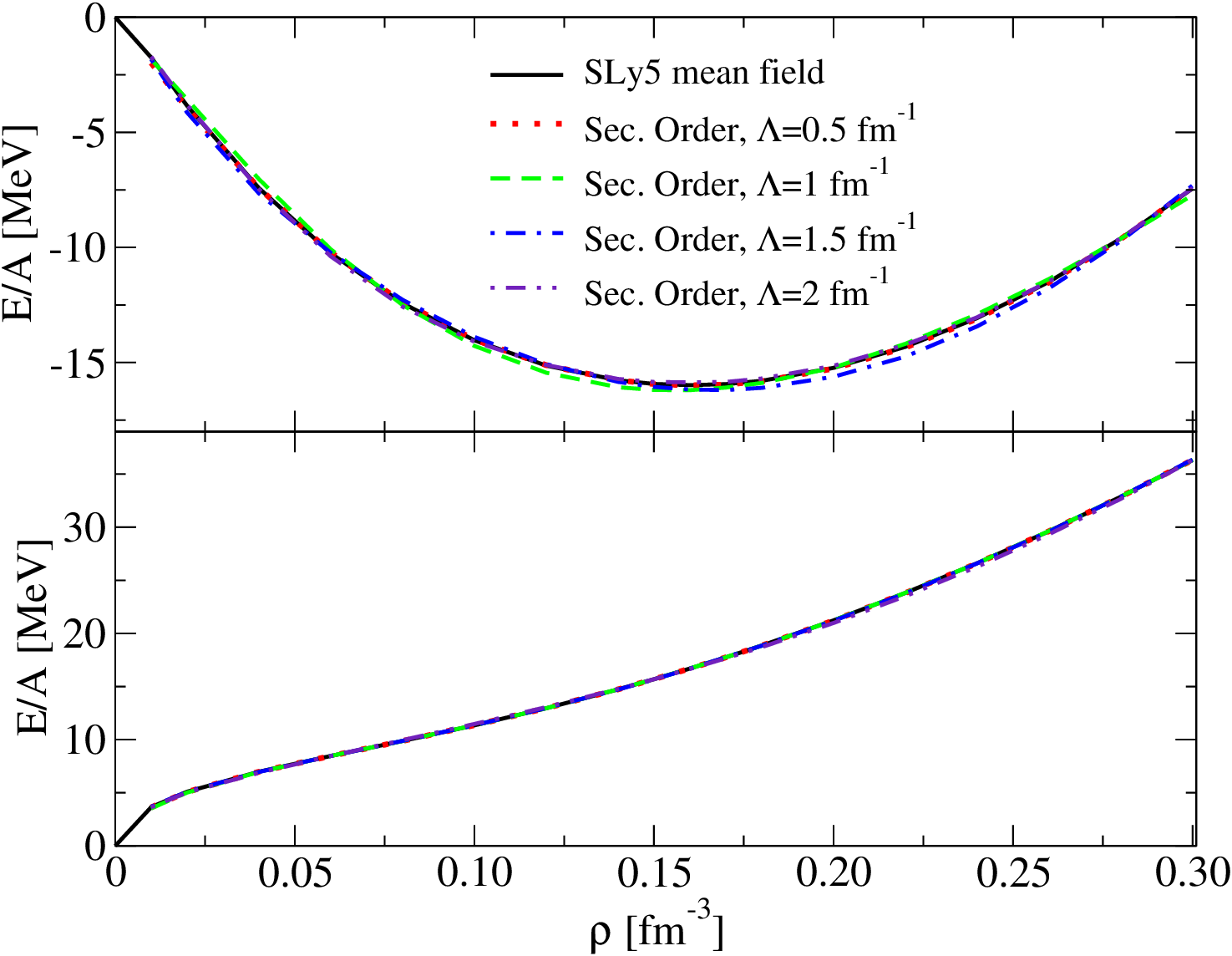}
\caption{(Color online) Second--order EOS's for symmetric (a)  and pure neutron matter (b) adjusted on the SLy5--mean--field EOS's, with an effective mass equal to the 
mean--field effective mass and with the inclusion of the rearrangement terms, for different values of the cutoff.}
\label{mstarrearraglobal}
\end{figure}

For the case of cutoff regularization, we present in Fig. \ref{mstarrearraglobal} the curves obtained with a global fit including symmetric and neutron matter (the corresponding parameters and $\chi^2$ can be found in Table VII). The incompressibility modulus ranges from 200 to 250 MeV, according to the different cutoff values and the 
quality of the fit is very good as indicated by the $\chi^2$ values.

The same global does not provide any good results for the case of dimensional regularization. 
We have then performed separately the two fits for symmetric and neutron matter (Figs. 28 and 29 and Tables VIII and IX). 
The incompressibility modulus is equal to 250 MeV. The quality of the fit 
is now less good but still reasonably good in the case of symmetric matter 
(average deviation of 13\%) and acceptable for neutron matter (average 
deviation of 21\%).
 
We see here the importance of including the density--dependent part of the Skyrme force. 
The inclusion of such part without the rearrangement terms allowed us in the previous section to shift the equilibrium point of symmetric matter to the correct one (compared to the EOS obtained in Ref. \cite{kaiser} where the $t_3$ part of the Skyrme force was totally neglected). The inclusion of the rearrangement terms allows us now to correctly describe the EOS of symmetric matter also in the other density regions and to have a reasonable value for the incompressibility modulus. 

\begin{figure}[tbp]
\includegraphics[scale=0.35]{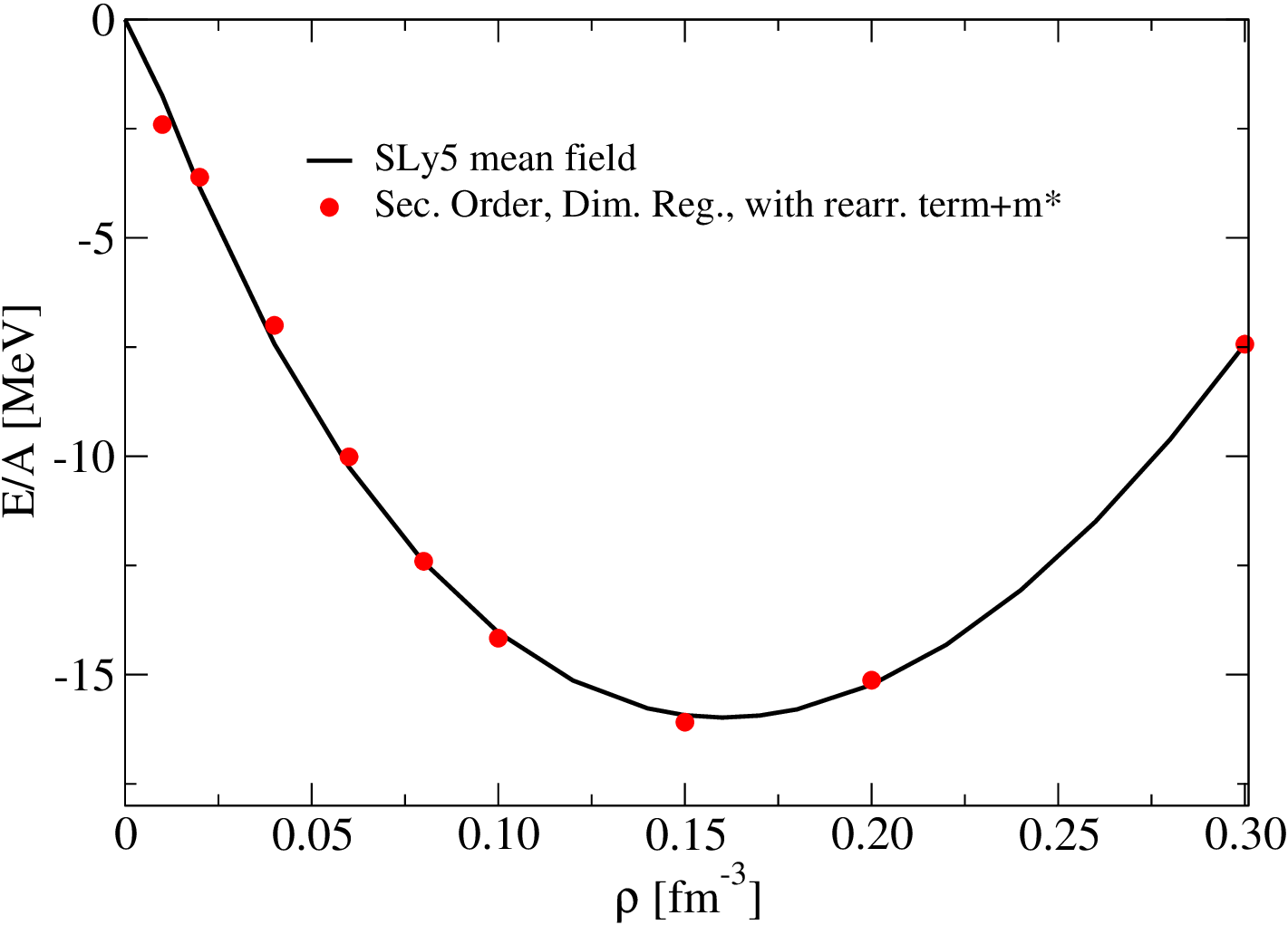}
\caption{(Color online) Same as in Fig. \ref{symmdrfit}, but with an effective mass equal to the 
mean--field effective mass and with the inclusion of the rearrangement terms.}
\label{mstarrearrds}
\end{figure}

\begin{figure}[tbp]
\includegraphics[scale=0.35]{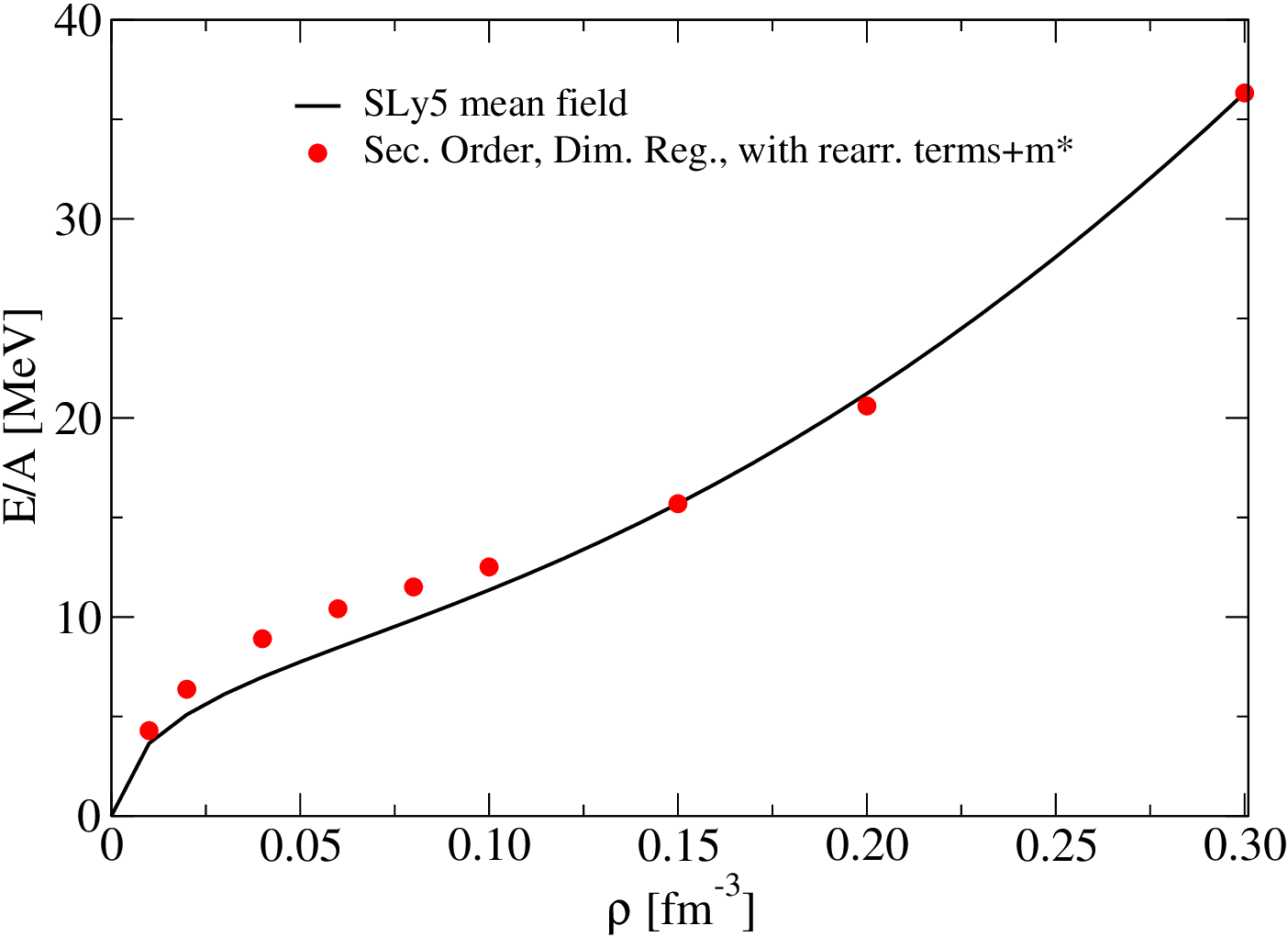}
\caption{(Color online) Same as in Fig. \ref{mstarrearrds}, but for neutron matter.}
\label{mstarrearrdsnm}
\end{figure}


\clearpage

\section{Conclusions}

We have presented a study devoted to the computation of the second--order correction in the EOS of symmetric, neutron, and asymmetric matter, with the use of the Skyrme effective interaction. Owing to the zero--range of such force, the second--order contribution to the nuclear--matter EOS diverges and a momentum cutoff must be used to regularize the divergent integrals. The  divergence is linear in the momentum cutoff in the simplified $t_0-t_3$ model \cite{moghraprl} and goes like the fifth power of the cutoff in the case where the velocity--dependent terms of the Skyrme interaction are also included. For simplicity, we have omitted the spin--orbit and tensor terms in the used expression of the Skyrme interaction. 
In addition to the occurrence of an ultraviolet divergence, second--order calculations performed  
with such an effective interaction present also a well--known risk of 
double counting, because the parameters of the force are adjusted to reproduce observables with leading--order (mean--field) calculations. 
This implies that the parameteres already contain in an implicit way some correlations. 
The cancellation of such double counting is then required. 

We have treated the ultraviolet divergence appearing at second order in the EOS of nuclear matter by using cutoff and dimensional regularizations, as was done in Refs. \cite{moghra2012-1} and \cite{moghra2012-2}, respectively: in the first case, the full second--order correction is calculated with all the cutoff--dependent terms. The analitycal derivation of all the terms is presented for 
symmetric and neutron matter. Results obtained numerically with a Monte Carlo integration for symmetric, neutron, and asymmetric matter are discussed. For each value of the introduced momentum cutoff, a new set of parameters is obtained by adjusting the second--order EOS to the chosen benchmark SLy5--mean--field EOS. This procedure eliminates both double--counting problems and divergences. In the case of dimensional regularization, only the finite part of the EOS is kept. In such case, only double--counting problems arise and they are removed, also this time, by an adjustment of the parameters. Unique sets of parameters are produced for each type of EOS (no cutoff). 
The objective of this work is to present in a complete and detailed form revised results and figures with respect to those illustrated in Refs. \cite{moghra2012-1} and \cite{moghra2012-2}. We have realized recently that those results are incomplete in some aspects concerning the analytical derivation of specific second--order contributions.   
In addition, we have evaluated the effects associated to the rearrangemet terms entering in second--order calculations and related to the density dependence of the interaction. Sets of parameters adjusted for symmetric and neutron matter, taking into account an effective mass evaluated at the mean--field level, and including also the proper rearrangement terms are provided. 
Such sets  may be considered a very reasonable starting point within our general objective, that is 
the construction of generalized effective interactions that are specially designed to be used in a beyond--mean--field scheme: parameters 
are adjusted at the same level of the performed calculations (no double counting) and 
produce results which are independent of the chosen energy or momentum cutoff. Such regularized interactions will open the possibility of performing robust beyond--mean--field applications to finite nuclei. 

\section*{Acknowledgments}
G. C. acknowledges discussions with N. Kaiser and M.G. aknowledges discussions with A. Pastore. 

\begin{widetext}

\begin{table}
\centering
\caption{Parameter sets obtained with the fit of the second--order EOS of symmetric matter for different values of the cutoff $\Lambda$ compared with the original set SLy5. In the last column, the $\chi^2$ values are shown. }\vspace{0.5cm}\label{symmetric_parameters}
\begin{tabular}{  c c c c c c c c c c c }
    \hline
    \hline
  &    $\quad t_0$ & $t_1$ & $t_2$ & $t_3$ & $x_0$ & $x_1$ & $x_2$ & $x_3$ & $\alpha$ & $\chi^2$  \\
 & (MeV fm$^3$) & (MeV fm$^5$) & (MeV fm$^5$) & (MeV fm$^{3+3\alpha}$) & & & & &  & \\
SLy5 &-2484.88    &    483.13   &   -549.40   & 13763.0&0.778     &    -0.328    &  -1.0 &1.267   &   0.16667 & \\
\hline
  $\Lambda$(fm$^{-1})$ & &  &  &  &  &  &  &  &    \\
 0.5 & -1461.868  &  497.986 & -1471.462 & 9915.064   &   0.5360 & -1.529 & -1.068  & 8.298   &   0.3201 & 2.08 10$^{-3}$\\
1.0 &  -1207.550  &      645.148  &     -1361.666 & 6942.115 &0.4854     & -2.106 &      -1.092  &      5.575    &  0.2831  &  1.46 10$^{-3}$\\
1.5 &  -1124.277    &    614.238    &  -1063.666 & 5711.048 & 0.4333     & -2.690   &  -1.349    &    5.103   &    0.2070   &   2.26 10$^{-2}$   \\
2.0 &  -530.285 &     227.301   &    -389.370 & 8927.183 & 0.9383     & -0.7346  &   -0.3959  &     0.4783   &    0.6680 & 3.13 10$^{-2}$
 \\
   \hline\hline
  \end{tabular}
  \end{table}

\begin{table}
\centering
\caption{Parameter sets obtained with the fit of the second--order EOS of neutron matter for different values of the cutoff $\Lambda$ compared with the original set SLy5. The $\chi^2$ values are shown in  the last column. }\vspace{0.5cm}\label{pure_parameters}
\begin{tabular}{  c c c c c c c c c c c }
    \hline
    \hline
   & $\quad t_0$ & $t_1$ & $t_2$ & $t_3$ & $x_0$ & $x_1$ & $x_2$ & $x_3$ & $\alpha$ & $\chi^2$  \\
 & (MeV fm$^3$) & (MeV fm$^5$) & (MeV fm$^5$) & (MeV fm$^{3+3\alpha}$) & & & & &  & \\
SLy5 &-2484.88    &483.13     &  -549.40    &   13736.0    &   0.778    &       -0.328    &     -1.0    &   1.267&    0.16667  & \\
\hline
$\Lambda$(fm$^{-1})$ & & & & & & & & & &   \\
0.5 &   -1859.971  &      474.575     &  -715.911& 16306.843 &      0.7102     & -0.7273   &  -0.9174    &    1.230   &    0.1625  &   2.19 10$^{-2}$ \\
1.0 &  -3524.393   &    569.934  &    -686.914 &11329.194& 1.048     & -0.7025  &    -0.8657    &    1.680   &    0.08796  & 2.14 10$^{-2}$\\
1.5 &  -3924.552     &   677.302     &  -724.123   &     11822.333& 1.398     & -0.5270   &   -0.9360    &    2.141 &0.09518 &  3.01 10$^{-2}$ \\
2.0 &-1701.409   &    159.406   &   -733.505 &24941.293& 1.246& -0.6726     & -1.1089   &    0.9313  &      0.09610  & 1.71 10$^{-2}$\\
   \hline\hline
  \end{tabular}
  \end{table}

\begin{table}
\centering
\caption{Parameter sets obtained with the fit of the second--order EOS of asymmetric matter ($\delta=0.5$) matter for different values of the cutoff $\Lambda$ compared with the original set SLy5. The $\chi^2$ values are shown in  the last column. }\vspace{0.5cm}\label{05_parameters}
\begin{tabular}{  c c c c c c c c c c c}
    \hline
    \hline
  &   $\quad t_0$ & $t_1$ & $t_2$ & $t_3$ & $x_0$ & $x_1$ & $x_2$ & $x_3$ & $\alpha$ & $\chi^2$  \\
 & (MeV fm$^3$) & (MeV fm$^5$) & (MeV fm$^5$) & (MeV fm$^{3+3\alpha}$) & & & & &  & \\
SLy5 & -2484.88    &483.13    &   -549.40    & 13763.0  &    0.778    &    -0.328    &  -1.0           &   1.267& 0.16667 &  \\
\hline
$\Lambda$(fm$^{-1}$) & & & & & & & & & & \\
0.5 &  -1639.753 &      584.294  &     -732.339   &     19750.210      & 0.5989  &  -0.339  &   -0.886      &  3.328& 0.284    & 4.76 10$^{-3}$
    \\
1.0 &  -1252.011  &     558.930   &   -744.859  &      10167.053      & 0.5887 &     -0.4652   &  -1.018   &     2.853   &    0.2272  & 1.19 10$^{-2}$ \\
1.5 &  -1328.894   &    594.658   &    -443.008& 8579.156& 0.4993   & -0.8051 &      -1.038   &     2.222   &    0.1640 & 1.76 10$^{-1}$ \\
2.0 &-745.086  &      358.107    &  -338.407 &9905.0844  &     0.3919    & -0.4444   &   -0.6662    &  0.6221  &     0.9432 & 6.87 10$^{-2}$  \\
   \hline\hline
  \end{tabular}
  \end{table}

\begin{table}
\centering
\caption{Parameter sets obtained with the global fit of the second--order EOS including symmetric, $\delta=0.5$ and neutron matter for different values of the cutoff $\Lambda$ 
compared with the original set SLy5. The $\chi^2$ values are shown in  the last column.  }\vspace{0.5cm}\label{gen_parameters}
\begin{tabular}{  c c c c c c c c c c c  }
    \hline
    \hline
   & $\quad t_0$ & $t_1$ & $t_2$ & $t_3$ & $x_0$ & $x_1$ & $x_2$ & $x_3$ & $\alpha$ & $\chi^2$  \\
 & (MeV fm$^3$) & (MeV fm$^5$) & (MeV fm$^5$) & (MeV fm$^{3+3\alpha}$) & & & & & &  \\
SLy5 &-2484.88  &483.13     &  -549.40  &   13736.0    &   0.778        &       -0.328  &     -1.0      &   1.267&      0.16667  & \\
\hline
$\Lambda$(fm$^{-1})$ & & & & & & & & & &  \\
0.5 &   -2245.402  &      493.322     &  -1832.783& 11961.86 &      0.7462     & -0.3936   &  -0.9684    &    1.309   &    0.1832     & 0.25\\
1.0 &  -1239.909   &    674.272  &    -387.948 &4687.107& 0.3649     & -0.5993  &    -1.1349    &    3.4299   &    0.5558  & 3.96 \\
1.5 &  -803.325     &   670.917     &  -42.426   &   4854.284& 0.1165     & -1.1436   &   -2.6727    &    3.4271 &1.1831  & 13.9 \\
2.0 &-668.075   &    80.904   &   0.8980 &8779.939& 0.1605& 0.3874     & -0.2652   &    0.0004687  &      1.4723  & 10.7 \\
   \hline\hline
  \end{tabular}
  \end{table}

\begin{table}
\centering
\caption{Parameter sets obtained with the fit of the dimensional--regularized second--order EOS compared with the original set SLy5. }\vspace{0.5cm}\label{dim_parameters}
\begin{tabular}{ c c c c c c c c c c c}
    \hline
    \hline
  &   $\quad t_0$ & $t_1$ & $t_2$ & $t_3$ & $x_0$ & $x_1$ & $x_2$ & $x_3$ & $\alpha$   \\
 & (MeV fm$^3$) & (MeV fm$^5$) & (MeV fm$^5$) & (MeV fm$^{3+3\alpha}$) & & & & &  \\
SLy5 & -2484.88 &483.13    &   -549.40  & 13763.0  &    0.778   &    -0.328     &  -1.0        &   1.267& 0.16667  & \\
\hline
$\delta$  & & & & & & & & &  \\
1 &  -3746.7 &   264.38  &     1607.4   &     -4537.9     & 0.8322  &  -1.3524  &   -1.1643      &  -13.7421& 2.0301
    \\
0.5 &  -920.60 &  544.55  &     -783.28   &   -879958      & 0.0289  &  -0.2788  &   -0.0681      &  -10650& 9.1666
    \\
0  & -938.36    &  975.87  &   -887.06     &  -348964.5   & -0.156  & -0.331  &   0.00265  & -0.442   &  3.104 

    \\
   \hline\hline
  \end{tabular}
  \end{table} 

\begin{table}
\centering
\caption{Parameter sets obtained with the global fit of the second--order EOS including symmetric and neutron matter for different values of the cutoff $\Lambda$
compared with the original set SLy5. Here the mean field effective mass is used in the second--order calculation. The $\chi^2$ values are shown in  the last column.  }\vspace{0.5cm}\label{gen_parameters_m}
\begin{tabular}{  c c c c c c c c c c c  }
    \hline
    \hline
   & $\quad t_0$ & $t_1$ & $t_2$ & $t_3$ & $x_0$ & $x_1$ & $x_2$ & $x_3$ & $\alpha$ & $\chi^2$  \\
 & (MeV fm$^3$) & (MeV fm$^5$) & (MeV fm$^5$) & (MeV fm$^{3+3\alpha}$) & & & & & &  \\
SLy5 &-2484.88  &483.13     &  -549.40  &   13736.0    &   0.778        &       -0.328  &     -1.0      &   1.267&      0.16667  & \\
\hline
$\Lambda$(fm$^{-1})$ & & & & & & & & & &  \\
0.5 &   -2254.55  &     555.99     &  -496.33& 12099.62 &      0.7429    & -0.3723   &  -0.9236    &    1.2982   &    0.1844    & 0.39\\
1.0 &  -1090.87   &   290.62 &    -552.05 &11613.08& 0.2255    & -5.3624  &    -0.8087    &    2.1436   &    0.6404  & 10 \\
1.5 &  -433.47     &   203.22     &  -280.96  &  6704.25& 0.3570     & -5.4828   &   -0.9308    &    2.7269&0.8262 &1.9 \\
2.0 &-667.98   &   79.28   &  35.51 &8842.99& 0.6230& 4.2801     & -0.2024  &    -0.3620 &      1.3769  & 1.4 \\
   \hline\hline
\label{aaa}
  \end{tabular}
  \end{table}

\begin{table}
\centering
\caption{Same as in Table \ref{gen_parameters_m}, with also the inclusion of rearrangement terms.   }\vspace{0.5cm}\label{aaa}
\begin{tabular}{  c c c c c c c c c c c  }
    \hline
    \hline
   & $\quad t_0$ & $t_1$ & $t_2$ & $t_3$ & $x_0$ & $x_1$ & $x_2$ &
$x_3$ & $\alpha$ & $\chi^2$  \\
 & (MeV fm$^3$) & (MeV fm$^5$) & (MeV fm$^5$) & (MeV fm$^{3+3\alpha}$)
& & & & & &  \\
SLy5 &-2484.88  &483.13     &  -549.40  & 13736.0    &   0.778
 &       -0.328  &     -1.0      &   1.267& 0.16667  & \\
\hline
$\Lambda$(fm$^{-1})$ & & & & & & & & & &  \\
0.5 &   -2226.39  &    833.57     &  -1054.76& 12615.02 &      0.7243
 & -0.3572   &  -0.8483    &    1.2719   & 0.2020    & 0.03\\
1.0 &  -9.43   &   1004.59 &    -3683.79 & -10057.96& -21.7819    &
3.1346  &    -1.2868    &  -0.2921   &    0.8324 & 3.48 \\
1.5 &  520.74     &   664.84     & -2226.01  & -8415.55 & 1.7992     &
-1.6278  &   -1.3517    &  -6.5449& 0.6387 & 2.15\\
2.0 & 894.01   &  -122.55  &  -968.37 & -3284.51& 1.3452& 8.1176     &
-1.9424 &    -16.1373 &   0.7411  & 2.20 \\
   \hline\hline
  \end{tabular}
  \end{table}


\begin{table}
\centering
\caption{Parameter set obtained for symmetric matter with the fit of the dimensional--regularized second--order EOS compared with the original set SLy5. 
The mean--field effective mass and rearrangement terms are included. 
The $\chi^2$ value is shown in  the last column.}\vspace{0.5cm}\label{dim_parameters_rearr}
\begin{tabular}{ c c c c c c c c c c c}
    \hline
    \hline
  &   $\quad t_0$ & $t_1$ & $t_2$ & $t_3$ & $x_0$ & $x_1$ & $x_2$ & $x_3$ & $\alpha$   & $\chi^2$\\
 & (MeV fm$^3$) & (MeV fm$^5$) & (MeV fm$^5$) & (MeV fm$^{3+3\alpha}$) & & & & &  & \\
SLy5 & -2484.88 &483.13    &   -549.40  & 13763.0  &    0.778   &    -0.328     &  -1.0        &   1.267& 0.16667  & \\
\hline
  & -1425.43 & -16732.70  &   1345.58      &  373005.96 &  -0.0279 &   -0.1885&   -5.2091&   -0.2014&    0.5796 & 184
    \\
   \hline\hline
  \end{tabular}
  \end{table} 

\begin{table}
\centering
\caption{ Same as in Table VIII but for neutron matter.}\vspace{0.5cm}
\begin{tabular}{ c c c c c c c c c c c}
    \hline
    \hline
  &   $\quad t_0$ & $t_1$ & $t_2$ & $t_3$ & $x_0$ & $x_1$ & $x_2$ & $x_3$ & $\alpha$   & $\chi^2$\\
 & (MeV fm$^3$) & (MeV fm$^5$) & (MeV fm$^5$) & (MeV fm$^{3+3\alpha}$) & & & & &  & \\
SLy5 & -2484.88 &483.13    &   -549.40  & 13763.0  &    0.778   &    -0.328     &  -1.0        &   1.267& 0.16667  & \\
\hline
& 1591.32 &   -837.21  &     -1498.53   &    2582.80      & 1.5534  &  1.7740  &   -0.9306      &  7.2733& 1.3874   &  466
    \\
   \hline\hline
  \end{tabular}
  \end{table}

\end{widetext}

\clearpage
\appendix

\section{Factors in front of the integrals}

\label{appendixa}

The second--order correction written in the proton (p) and neutron (n) basis is the
sum of nn, pp, and np contributions, that is 
\begin{eqnarray}
\Delta E^{(2)} &=&\frac{1}{2}\sum_{ij}\langle ij|VGV|ij\rangle _{pp}+\frac{1%
}{2}\sum_{ij}\langle ij|VGV|ij\rangle _{nn}  \notag \\
&&+\sum_{ij}\langle ij|VGV|ij\rangle _{np}.
\end{eqnarray}

\bigskip Here $i$ and $j$ are the labels of the two particles. Note that $%
\sum_{ij}=\frac{\Omega ^{3}}{(2\pi )^{9}}\int d^{3}\mathbf{k}_{1}\int d^{3}%
\mathbf{k}_{2}\int d^{3}\mathbf{q}$, and $V=v/\Omega $. For the np case, we
associate $i$ to n and $j$ to p. The factor 2 in the np part with respect to
nn and pp part accounts for the symmetric pn contribution.

When evaluating the matrix element, the exchange term is included by
inserting $(1-P_{x}P_{\sigma }P_{\tau })$ from the left, where $%
P_{x},P_{\sigma },P_{\tau }$ are the space, spin, and isospin exchange
operators, respectively. The exchange contribution is thus equal to the
direct one in the nn and pp channels in the cases of even--even and odd--odd
mixing of the interaction. In the cases of even--odd mixing of the
interaction, one has $1-P_{x}P_{\sigma }P_{\tau }= 1+P_{\sigma }P_{\tau } $
for the nn and pp channels ($P_x$ provides a minus sign). The exchange term
is always equal to zero in the np channel.

\subsection{$t_{0}^{2}$\textbf{\ (}$t_{3}^{2}$\textbf{) matrix element:}}

We start with the square of the term $v=t_{0}(1+x_{0}P_{\sigma })$
(analogous expressions may be written for the square of the $t_3$ term). 

(a) nn channel. In this case, $T=1$. The mixing of the interaction is
even--even ($l=0$) and this leads to $S=0$. $P_{\sigma}$ acting on the spin
singlet state provides a minus sign. One has

\begin{eqnarray}
&& \frac{1}{2} \sum_{ij}\langle ij|VGV|ij\rangle _{nn}  \notag \\
&=&\frac{t_{0}^{2}\Omega}{(2\pi )^{9}}\int d^{3}\mathbf{k}_{1}\int d^{3}%
\mathbf{k}_{2}\int d^{3}\mathbf{q}G\mathbf{[}\sum_{SM_S}\langle X^S_{M_S}
|(1+2x_{0}P_{\sigma }+x_{0}^{2})|X^S_{M_S}\rangle \mathbf{],}
\end{eqnarray}
where $M_S$ is the spin projection. By applying the spin exchange operator
and by performing the sum (there is only one term: $S=0$; $M_S=0$) one has 
\begin{eqnarray}
&&\frac{1}{2}\sum_{ij}\langle ij|VGV|ij\rangle _{nn}  \notag \\
&=&\frac{t_{0}^{2}\Omega}{(2\pi )^{9}}\mathbf{[}1-2x_{0}+x_{0}^{2}\mathbf{]}%
\int d^{3}\mathbf{k}_{1}\int d^{3}\mathbf{k}_{2}\int d^{3}\mathbf{q}G
\end{eqnarray}%
We introduce the quantities%
\begin{eqnarray}
\overline{\mathbf{k}}_{1} &=&\frac{\mathbf{k}_{1}}{k_{N}}, \; \overline{%
\mathbf{k}}_{2} =\frac{\mathbf{k}_{2}}{k_{N}}, \; \overline{\mathbf{q}} =%
\frac{\mathbf{q}}{k_{N}},  \label{bar} \\
\overline{\mathbf{k}} &=&\frac{\mathbf{k}}{k_{N}}, \; \overline{\mathbf{%
k^{\prime}}} =\frac{\mathbf{k^{\prime}}}{k_{N}},  \notag \\
\overline{\Lambda} &=&\frac{\Lambda}{k_{N}}, \; \overline{G} =G k_{N}^2. 
\notag
\end{eqnarray}%
Then:%
\begin{eqnarray}
&&\frac{1}{2}\sum_{ij}\langle ij|VGV|ij\rangle _{nn}  \notag \\
&=&\frac{t_{0}^{2}\Omega }{(2\pi )^{9}}\mathbf{[}1-2x_{0}+x_{0}^{2}\mathbf{]}%
k_{F_N}^{7}\int_{0}^{1}d^{3}\overline{\mathbf{k}}_{1}\int_{0}^{1}d^{3}%
\overline{\mathbf{k}}_{2}\int_{0}^{\overline{\Lambda} }d^{3}\overline{%
\mathbf{q}}\overline{G}.  \label{t0s}
\end{eqnarray}

(b) pp channel:

The pp contribution is the same as the nn contribution, with $%
k_{F_N}\rightarrow k_{F_P}$ in Eq.~(\ref{t0s}).

\bigskip

(c) np channel (in the np channel, the operator $P_{\sigma}$ provides a plus
sign when applied to the spin triplet states and a minus sign when applied
to the spin singlet state).

We introduce the quantity 
\begin{equation}
a=\frac{k_{F_P}}{k_{F_N}}=\left(\frac{1-\delta }{1+\delta }\right)^{1/3}.
\label{a}
\end{equation}%
Then:

\begin{eqnarray}
&&\sum_{ij}\langle ij|VGV|ij\rangle _{np}  \notag \\
&=&\frac{t_{0}^{2}\Omega }{(2\pi )^{9}}k_{F_N}^{7}\int_{0}^{1}d^{3}\overline{%
\mathbf{k}}_{1}\int_{0}^{a}d^{3}\overline{\mathbf{k}}_{2}\int_{0}^{\overline{%
\Lambda} }d^{3}\overline{\mathbf{q}}\overline{G}\mathbf{[}\sum_{S
M_S}\langle X^S_{M_S} |(1+2x_{0}P_{\sigma }+x_{0}^{2})|X^S_{M_S}\rangle 
\mathbf{]} \\
&=&4\frac{t_{0}^{2}\Omega }{(2\pi )^{9}}k_{F_N}^{7}(1+x_{0}+x_{0}^{2})%
\int_{0}^{1}d^{3}\overline{\mathbf{k}}_{1}\int_{0}^{a}d^{3}\overline{\mathbf{%
k}}_{2}\int_{0}^{\overline{\Lambda} }d^{3}\overline{\mathbf{q}}\overline{G}.
\end{eqnarray}

\bigskip

\subsection{$t_{0}(t_3)t_{2}$\textbf{\ part:}}

This contribution is equal to zero in the case of symmetric and neutron
matter, because it mixes even and odd terms of the interaction. One has 
\begin{equation}
vGv=t_{0}(1+x_{0}P_{\sigma })t_{2}(1+x_{2}P_{\sigma })\mathbf{k}^{^{\prime }}%
\mathbf{\cdot k}G.
\end{equation}%
By including the direct and exchange terms, 
\begin{eqnarray}
\langle ij|vGv|ij\rangle = \mathbf{k}^{^{\prime }}\mathbf{\cdot k}G\langle
ij|t_{0}t_{2}(1+x_{2}P_{\sigma }+x_{0}P_{\sigma }+x_{0}x_{2})(1+P_{\sigma
}P_{\tau })|ij\rangle.
\end{eqnarray}%
%
%
%

(a) nn channel:

In this case, both particles are neutrons and the antisymmetrization
condition must be imposed. Therefore, as in Sect. II.A, the odd--even mixing
of the interaction is not allowed. Thus,%
\begin{equation*}
\sum_{ij}\langle ij|VGV|ij\rangle _{nn}=0.
\end{equation*}%
%
%
%

(b) pp channel:

The pp part is the same as the nn contribution and thus equal to zero.

\bigskip

(c) np channel:

Now the two particles are not identical if their densities are different
(different Fermi momenta). We have the following non-vanishing term:

\begin{eqnarray}
&&\sum_{ij}\langle ij|VGV|ij\rangle _{np}  \notag \\
&=&\frac{t_{0}t_{2}\Omega }{(2\pi )^{9}}k_{F_N}^{9}\int_{0}^{1}d^{3}\overline{%
\mathbf{k}}_{1}\int_{0}^{a}d^{3}\overline{\mathbf{k}}_{2}\int_{0}^{\overline{%
\Lambda }}d^{3}\overline{\mathbf{q}}\overline{\mathbf{k}}^{^{\prime }}%
\mathbf{\cdot \overline{k}}\overline{G}\mathbf{[}\sum_{SM_{S}}\langle
X_{M_{S}}^{S}|1+(x_{0}+x_{2})P_{\sigma }+x_{0}x_{2}|X_{M_{S}}^{S}\rangle 
\mathbf{]}  \notag \\
&=&2\frac{t_{0}t_{2}\Omega }{(2\pi )^{9}}%
k_{F_N}^{9}(2+x_{0}+x_{2}+2x_{0}x_{2})\int_{0}^{1}d^{3}\overline{\mathbf{k}}%
_{1}\int_{0}^{a}d^{3}\overline{\mathbf{k}}_{2}\int_{0}^{\overline{\Lambda }%
}d^{3}\overline{\mathbf{q}}\overline{\mathbf{k}}^{^{\prime }}\mathbf{\cdot 
\overline{k}}\overline{G}.  \label{t02np}
\end{eqnarray}%
Note that if protons and neutrons have the same density, then they can be
considered as identical particles. This is reflected in the above equation:
by setting $a=1$, the integral leads to zero, as for the nn and pp cases.

All the other matrix elements can be evaluated in the same way as above. We
list the results of the other terms below.

\subsection{$t_{0}(t_3)t_{1}$\textbf{\ part:}}

(a) nn channel:%
\begin{eqnarray}
&&\frac{1}{2}\sum_{ij}\langle ij|VGV|ij\rangle _{nn}  \notag \\
&=&\frac{1}{2}\frac{t_{0}t_{1}\Omega}{(2\pi )^{9}}%
k_{F_N}^{9}(1-x_{0}-x_{1}+x_{0}x_{1})\int_{0}^{1}d^{3}\overline{\mathbf{k}}%
_{1}\int_{0}^{1}d^{3}\overline{\mathbf{k}}_{2}\int_{0}^{\overline{\Lambda}
}d^{3}\overline{\mathbf{q}}(\overline{\mathbf{k}}^{\prime 2}+\overline{%
\mathbf{k}}^{2})\overline{G}  \label{t01}
\end{eqnarray}

(b) pp channel:

The pp part is the same as the nn part, by replacing $k_{F_N}\rightarrow k_{F_P}$
in Eq.~(\ref{t01}).

\bigskip

(c) np channel:

\begin{eqnarray}
&&\sum_{ij}\langle ij|VGV|ij\rangle _{np}  \notag \\
&=&\frac{t_{0}t_{1}\Omega}{(2\pi )^{9}}k_{F_N}^{9}(2+x_{0}+x_{1}+2x_{0}x_{1})%
\int_{0}^{1}d^{3}\overline{\mathbf{k}}_{1}\int_{0}^{a}d^{3}\overline{\mathbf{%
k}}_{2}\int_{0}^{\overline{\Lambda} }d^{3}\overline{\mathbf{q}}(\overline{%
\mathbf{k}}^{\prime 2}+\overline{\mathbf{k}}^{2})\overline{G}.
\end{eqnarray}

\bigskip

\subsection{$t_{1}t_{2}$\textbf{\ part:}}

(a) nn channel. It is analogous to the nn part of the $t_0t_2$
part(even--odd mixing).

\begin{eqnarray}
&&\sum_{ij}\langle ij|VGV|ij\rangle _{nn}=0
\end{eqnarray}

(b) pp channel:

The pp part is the same as the nn part ans is 
equal to zero.

\bigskip

(c) np channel:

\begin{eqnarray}
&&\sum_{ij}\langle ij|VGV|ij\rangle _{np}  \notag \\
&=&\frac{t_{1}t_{2}\Omega}{(2\pi )^{9}}k_{F_N}^{11}(2+x_{1}+x_{2}+2x_{1}x_{2})%
\int_{0}^{1}d^{3}\overline{\mathbf{k}}_{1}\int_{0}^{a}d^{3}\overline{\mathbf{%
k}}_{2}\int_{0}^{\overline{\Lambda} }d^{3}\overline{\mathbf{q}}\mathbf{(%
\overline{\mathbf{k}}^{\prime 2}+\overline{\mathbf{k}}^{2})\overline{k}}%
^{^{\prime }}\mathbf{\cdot \overline{k}}\overline{G}.
\end{eqnarray}

\subsection{$t_{1}^{2}$\textbf{\ part:}}

(a) nn channel: 
\begin{eqnarray}
&&\frac{1}{2}\sum_{ij}\langle ij|VGV|ij\rangle _{nn}  \notag \\
&=&\frac{1}{4}\frac{t_{1}^{2}\Omega}{(2\pi )^{9}}\mathbf{[}1-2x_{1}+x_{1}^{2}%
\mathbf{]}k_{F_N}^{11}\int_{0}^{1}d^{3}\overline{\mathbf{k}}%
_{1}\int_{0}^{1}d^{3}\overline{\mathbf{k}}_{2}\int_{0}^{\overline{\Lambda}
}d^{3}\overline{\mathbf{q}}\mathbf{(\overline{\mathbf{k}}^{\prime 2}+%
\overline{\mathbf{k}}^{2})}^{2}\overline{G},
\end{eqnarray}

(b) pp channel:

The pp part is the same as the nn part, with $k_{F_N}\rightarrow k_{F_P}$.

\bigskip

(c) np channel:

\begin{eqnarray}
&&\sum_{ij}\langle ij|VGV|ij\rangle _{np}  \notag \\
&=&\frac{t_{1}^{2}\Omega }{(2\pi )^{9}}k_{F_N}^{11}(1+x_{1}+x_{1}^{2})%
\int_{0}^{1}d^{3}\overline{\mathbf{k}}_{1}\int_{0}^{a}d^{3}\overline{\mathbf{%
k}}_{2}\int_{0}^{\overline{\Lambda} }d^{3}\overline{\mathbf{q}}\mathbf{(%
\overline{\mathbf{k}}^{\prime 2}+\overline{\mathbf{k}}^{2})}^{2}\overline{G}.
\end{eqnarray}

\subsection{$t_{2}^{2}$\textbf{\ part:}}

(a) nn channel. In this case, $T=1$. The mixing of the interaction is
odd--odd ($l=1$) and this leads to $S=1$. $P_{\sigma}$ provides a plus sign
for each of the triplet states: 
\begin{eqnarray}
&&\frac{1}{2}\sum_{ij}\langle ij|VGV|ij\rangle _{nn}  \notag \\
&=&3\frac{t_{2}^{2}\Omega }{(2\pi )^{9}}\mathbf{[}1+2x_{2}+x_{2}^{2}\mathbf{]%
}k_{F_N}^{11}\int_{0}^{1}d^{3}\overline{\mathbf{k}}_{1}\int_{0}^{1}d^{3}%
\overline{\mathbf{k}}_{2}\int_{0}^{\overline{\Lambda} }d^{3}\overline{%
\mathbf{q}}\mathbf{(\overline{\mathbf{k}}^{^{\prime }}\mathbf{\cdot 
\overline{k}})}^{2}\overline{G},
\end{eqnarray}

(b) pp channel:

The pp part is the same as the nn part, with $k_{F_N}\rightarrow k_{F_P}$.

\bigskip

(c) np channel:

\begin{eqnarray}
&&\sum_{ij}\langle ij|VGV|ij\rangle _{np}  \notag \\
&=&4\frac{t_{2}^{2}\Omega }{(2\pi )^{9}}k_{F_N}^{11}(1+x_{2}+x_{2}^{2})%
\int_{0}^{1}d^{3}\overline{\mathbf{k}}_{1}\int_{0}^{a}d^{3}\overline{\mathbf{%
k}}_{2}\int_{0}^{\overline{\Lambda} }d^{3}\overline{\mathbf{q}}\mathbf{(%
\overline{\mathbf{k}}^{^{\prime }}\mathbf{\cdot \overline{k}})}^{2}\overline{%
G}.
\end{eqnarray}

\bigskip

\end{document}